\documentclass[journal]{IEEEtran}
\ifCLASSINFOpdf
\else
\fi
\usepackage{graphicx}
\usepackage{booktabs}
\usepackage{graphicx}
\usepackage{subfig}
\usepackage{url}
\usepackage{amsmath}
\usepackage{upgreek}
\usepackage{color}
\usepackage{algorithm}
\usepackage{algorithmicx}
\usepackage{algpseudocode}
\usepackage{amsmath,amssymb,bm}
\usepackage{empheq}
\usepackage{cases}


\newtheorem{theorem}{Theorem}

\newtheorem{corollary}{Corollary}

\newtheorem{lemma}{Lemma}

\newcommand{\vectw}[2]{\bm{#1}^{#2}}

\newcommand{\eg}{\emph{e.g.},}

\newcommand{\footurl}[1]{\footnote{\url{#1}}}

\newcommand{\changed}{}
\graphicspath{{include/}}

\begin{document}
\title{The Beauty of the Commons: Optimal Load Sharing by Base Station Hopping in Wireless Sensor Networks}
\author{Runwei~Zhang,
        Francois~Ingelrest,
        Guillermo~Barrenetxea,
        Patrick~Thiran,~\IEEEmembership{Fellow,~IEEE},
        and~Martin~Vetterli,~\IEEEmembership{Fellow,~IEEE,~ACM}
\thanks{R. Zhang, P. Thiran and M. Vetterli are with I\&C, \'{E}cole Polytechnique F\'{e}d\'{e}rale de Lausanne (EPFL), Lausanne, Switzerland (e-mail: runwei.zhang@epfl.ch, patrick.thiran@epfl.ch and martin.vetterli@epfl.ch)
}
\thanks{G. Barrenetxea is with Swisscom, Bern, Switzerland (e-mail: guillermo.barrenetxea@gmail.com ) }
\thanks{F. Ingelrest is with Sensorscope Sarl, Lausanne, Switzerland (e-mail: francois.ingelrest@sensorscope.ch).}
}
\maketitle
\begin{abstract}
In wireless sensor networks (WSNs), the base station (BS) is a critical sensor node whose failure causes severe data losses. 
Deploying multiple fixed BSs improves the robustness, yet requires all BSs to be installed with large batteries and large energy-harvesting devices due to the high energy consumption of BSs. 
In this paper, we propose a scheme to coordinate the multiple deployed BSs such that the energy supplies required by individual BSs can be substantially reduced. 
In this scheme, only one BS is selected to be active at a time and the other BSs act as regular sensor nodes. 
We first present the basic architecture of our system, including how we keep the network running with only one active BS and how we manage the handover of the role of the active BS. 
Then, we propose an algorithm for adaptively selecting the active BS under the spatial and temporal variations of energy resources. This algorithm is simple to implement but is also asymptotically optimal under mild conditions.
Finally, by running simulations and real experiments on an outdoor testbed, we verify that the proposed scheme is energy-efficient, has low communication overhead and {\changed reacts rapidly to network changes}.
\end{abstract}
\begin{IEEEkeywords}
Wireless sensor networks, multiple base station, load sharing, renewable energy supply
\end{IEEEkeywords}

\IEEEpeerreviewmaketitle

\section{Introduction}
\label{sec:introduction}
\IEEEPARstart{W}{ireless} sensor networks (WSNs) are composed of autonomous sensor nodes that monitor physical conditions. 
{Regular sensor nodes} in WSNs perform sensing and transmit the captured data to a \emph{base station} (BS) by using  \emph{short-range communication}, \textit{e.g.}, 802.15.4/Zigbee, in a multi-hop manner. 
The BS is the key sensor node that collects data across the WSN and then forwards it to a remote server by using \emph{long-range communication}, \textit{e.g.}, GSM/GPRS. 
It serves as a communication bridge between the sensing field and the remote server. {\changed Therefore, the BS is the bottleneck in a WSN: if some regular sensor nodes are disconnected from the BS, they will have data losses; if the BS fails, the whole network will get stuck.}

{\changed During the past few years, we have been working on the Sensorscope project~\cite{2010-TOSN-INGELREST}, whose objective is to deploy a WSN on the glaciers in the snow mountains to monitor the climate changes. Due to the harsh environment, the BS might fail and  the network might split into smaller networks due to connectivity problems. To let the network be {\emph{``robust''}}, or in other words, be able to recover from such incidents, we have to install multiple BSs in the sensing field, as many others do~\cite{Bogdanov2004575,4283889}. Because of the high energy consumption of long-range communication, all BSs are required to be equipped with large batteries and large solar panels. This is definitely undesirable because of the increased hardness of both deployment and maintenance.}

\begin{figure}[t]
    \centering
    \includegraphics[width=0.9\linewidth]{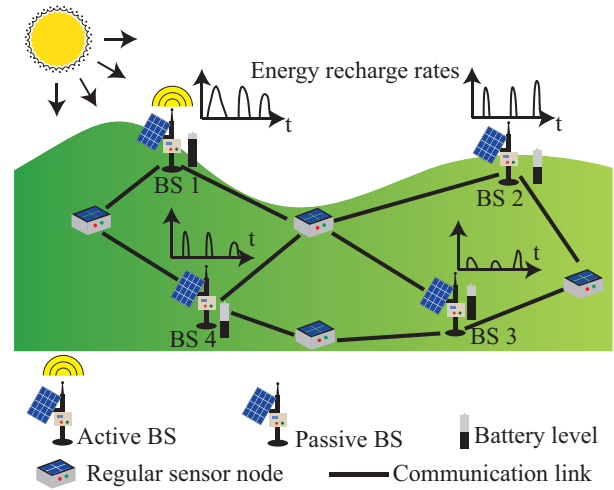}
    \caption{A WSN with the proposed scheme that deploys multiple BSs, keeps only one of them active and adaptively re-selects this active BS. At the current time, BS $1$ is active. 
Some time later, the active BS will be re-selected based on the states of the network, \textit{e.g.}, battery levels. 
By using this scheme, the temporally and spatially varying energy resources of all BSs are fully utilized.} 
    \label{fig:new-wsn}
\end{figure}

In this paper, we propose a novel scheme for coordinating the energy resources available to all the deployed BSs such that the sizes of energy sources for individual BSs can be substantially reduced.
The idea is to shut down unnecessary BSs and to keep only one active BS, as shown in Figure~\ref{fig:new-wsn}. To share the high load of being the active BS, we adaptively and iteratively select the BS that is activated. The active BS collects data and maintains long-range communication with the remote server. 
Meanwhile, passive BSs behave as regular sensor nodes. 
They turn off their long-range communication devices, only sampling and forwarding data by using short-range communication.
{\changed When the network has connectivity problems and splits into several connected components, the aforementioned active-BS selection process automatically takes place in all these small components.
In each connected component,} the high energy consumption of using long-range communication for the active BS is shared among all BSs. The batteries of all BSs form a pool, virtually resulting in a larger global power source. To build a sustainable WSN, the requirement is that the total energy harvested by all BSs sustains the consumption of the active BS. Consequently, the size of the individual power sources can be substantially reduced. 

{\changed Because the scheme for coordinating multiple BSs is unique}, we have to solve the following practical issues: (i) when the network is connected, how to start the WSN into the state with only one active BS, (ii) how to adaptively gather the information and decide the next active BS, (iii) how to manage the handover of the active BS {\changed and (iv) how to detect and recover from a network split or a failure of the active BS.} The solutions we provide to these issues are distributed and robust.

{\changed In each connected component of the network, we have to adaptively re-select the active BS. The first idea coming to one's mind is to use Round-Robin (RR):} we let all BSs be sequentially active with equal time. {\changed  However, RR is not necessarily optimal due to the following two reasons: (i) The energy recharged from solar panels of different BSs might be different because the solar panels might have different positions, different angles to the sun and different energy conversion efficiency. (ii) Passive BSs might have different energy consumptions due to different loads of short-range communication.}
We propose a simple but powerful algorithm, which we call \emph{``Highest Energy First''} (HEF), and which adaptively selects the BS with the highest available energy to be active. The appealing feature of HEF is that it requires little information as input and yet fits perfectly for the WSN paradigm. The active BS only needs to gather the battery levels of passive BSs. This algorithm is proved to be asymptotically optimal under mild conditions. 


To evaluate our  proposed scheme, we first run several simulations on the simulator Omnet++/Castalia~\cite{4351800} and next run real experiments on an outdoor testbed. Simulation results show that HEF is energy-efficient, has low communication overhead and {\changed reacts rapidly to network changes}. The real experiments lasted for $15$ days, and their results show that by using HEF to coordinate $3$ BSs, the lifetime of the WSN is prolonged by a factor of $3$ to $4$. The enhancement will be more pronounced if HEF is used on a larger number of cooperative BSs.

The main contributions of this paper are as follows:
\begin{enumerate}
\item We propose a novel scheme that deploys multiple BSs, keeps only one BS active at a time and adaptively re-selects the active BS. By using this scheme, the temporally and spatially varying energy resources available to all BSs are efficiently utilized, and therefore the energy supplies of individual BSs can be reduced substantially.

\item We propose an adaptive algorithm HEF for re-selecting the active BS. This algorithm requires little information exchange in the WSN and is easy to implement. We show that under certain mild conditions, this algorithm is asymptotically optimal. 

\item We discuss the implementation issues of HEF on real WSNs. In particular, we discuss how to start the network, how to gather the needed information and how to hand over the active BS. The solutions we provide are distributed and robust.
\item To evaluate the proposed scheme, we run simulations on the simulator Omnet++/Castalia and real experiments on an outdoor testbed. To the best of our knowledge, it is the first installation of a real testbed with multiple cooperative BSs. The obtained results show that our proposed scheme is energy-efficient, has low communication overhead and {\changed reacts rapidly to network changes}.
\end{enumerate}

The outline of this paper is as follows. First, we show related work in Section~\ref{sec:related_work}. Then, we formulate the adaptive BS selection problem in Section~\ref{sec:formulation}. Next, we propose the HEF algorithm and prove its asymptotic optimality in Section~\ref{sec:scheduling_algorithms}. In Section~\ref{sec:architecture}, we describe the implementation issues on real WSNs. We show results from simulations in Section~\ref{sec:simulations} and from experiments in Section~\ref{sec:real_experiment}. Finally, we conclude in Section~\ref{sec:conclusion}. 
 
\section{Related Work}
\label{sec:related_work}
This paper relates closely with the works on deploying multiple fixed BSs, the works on physically moving the BS and the works on energy management of energy harvesting WSNs.

{\bf{Deploying multiple fixed BSs}}: Researchers have previously proposed to deploy multiple fixed and always-active BSs for enhancing the robustness of WSNs and for reducing the energy consumption of short-range communication. Vincze~\textit{et al.}~\cite{4283889} optimize the locations of the multiple BSs in order to minimize the average distance from regular sensor nodes to BSs. Andrej \textit{et al.}~\cite{Bogdanov2004575} show that the problem of finding the optimal locations of BSs to maximize the sensing data rate under energy constraints is NP-hard. They propose a greedy heuristic to solve it. 
These works all implicitly assume that BSs have infinite energy supplies, which requires the installation of large batteries and large energy harvesting devices on all BSs. 

{\bf{Physically moving the BS}}: This paper is also inspired by previous works on physically moving the BSs. Their goals are usually to mitigate the energy hole problem caused by the high energy consumption of sensor nodes around the only BS. Optimizing the continuous travel path of the BS to maximize the lifetime of the WSN is usually hard. Bi \textit{et al.}~\cite{Bi2007} propose a simple strategy that intuitively moves the BS towards the nodes with high residual energy and away from the nodes with low residual energy.
Shi \textit{et al.}~\cite{Shi2011} reduce the infinite search space of the continuous travel path of the BS into a finite subset of discrete sites. They show that the simplification still guarantees the achieved network lifetime to be within $1-\epsilon$ of the maximum network lifetime, where $\epsilon$ can be set arbitrarily small. However, adding mobility to BSs is often infeasible, for example, in remote environmental monitoring applications~\cite{2008-IPSN-BARRENETXEA}. 

{\bf{Energy management of energy harvesting devices}}: The works in this area implicitly assume that the locations of the BSs are fixed and design energy spending policies of energy harvesting WSNs. They either assume that: (i) the exact profiles of energy recharge rates are deterministically known at the very beginning~\cite{Fan:2008:SFR:1460412.1460436,5461958},  (ii) the probability distributions of the energy recharge rates are known in advance~\cite{5934952}, or (iii) the probability distribution of the energy recharge rates are unknown but are assumed to be stationary in some sense, for example, i.i.d~\cite{5403539}. 
Our work falls into this third category, and we make a weaker assumption that the energy recharge rates have constant conditional expectations at all time.

In this paper, we set up multiple BSs for enhancing the robustness. {\changed  To efficiently use the available energy to all BSs, we adaptively re-select one active BS for using the long-range communication. We could go further by considering the scheme which adaptively re-selects multiple active BSs and jointly optimizes the available energy of both BSs and regular sensor nodes, but such a scheme will largely increase the implementation complexity and therefore is left for future work.}


\section{Design of the scheme}
\label{sec:architecture}
{\changed In this section, we discuss the practical issues for coordinating multiple BSs in a real WSN.} In particular, we tackle the following problems: (i) how the network starts into the state with only one active BS, (ii) how the active BS gathers the information needed for the selections, (iii) how the active BS hands over the role to the selected successor, and {\changed (iv) how the network recovers from unexpected failures.} Before discussing these issues, we first briefly review the overall architecture of the system.

\subsection{General Architecture}
In our architecture, time is partitioned into slots whose lengths $\tau$ are two hours each. At the beginning of each time slot, 
one active BS is selected in a distributed way. It begins broadcasting beacons and notifying the whole network. Upon receiving these beacons, passive BSs and regular sensor nodes update their routing tables and forward these beacons. Every sensor node takes sensing samples at a constant rate. The sensed data are then forwarded to the active BS by using short-range communication in a multi-hop fashion. The active BS collects all the data packets and forwards them to the remote server. In the next time slot, the active BS remains the same or it hands over the role to its successor, depending on the output result of HEF. Then, the new active BS starts broadcasting the beacons and the whole process is repeated.

In the following, we will show how the network manages synchronization, MAC protocols, routing protocols and the usage of GSM/GPRS. The interested reader can refer to our previously published work for more details and justifications of the choices~\cite{2010-TOSN-INGELREST}.


\subsubsection{Synchronization}

All sensor nodes are synchronized on \emph{Universal Coordinated Time} (UTC), retrieved by the active BSs when they connect to our server. The current time $T_c$ is inserted into beacons through MAC time-stamping~\cite{2003-SENSYS-GANERIWAL}. To estimate $T_c$, we use the crystal of sensor nodes to compute the elapsed time since the last update of  UTC. This mechanism, although simplistic, allows for a synchronization in the order of one millisecond, which is sufficient in our application.

\subsubsection{MAC protocols}
In the MAC layer, we adopt the commonly used S-MAC~\cite{1019408} and T-MAC~\cite{2003-SENSYS-VANDAM}. S-MAC is a collision avoidance MAC protocol with fixed duty-cycles for all sensor nodes. With T-MAC, sensor nodes dynamically adjust their duty cycles based on the communication loads.

\subsubsection{Routing protocols}

We use the gradient routing where sensor nodes send the data packets to their neighbors who have the shortest hop-distances to the active BS. We also make a few modifications on the classic gradient routing protocol, so that control messages for updating the active BS is specially handled, as will be discussed later.

\subsubsection{GSM/GPRS usage}

As the GSM/GPRS chip is an energy-hungry device (two orders of magnitude more than the short-range radio transceiver), its connection to the server is duty-cycled. There is an obvious trade-off between real-time information and energy savings. The typical connection interval that we use is $5$ minutes.

\subsection{Starting The Network}
\label{sec:bootstrap-process}
\begin{figure}
    \centering
    \includegraphics[width=\linewidth]{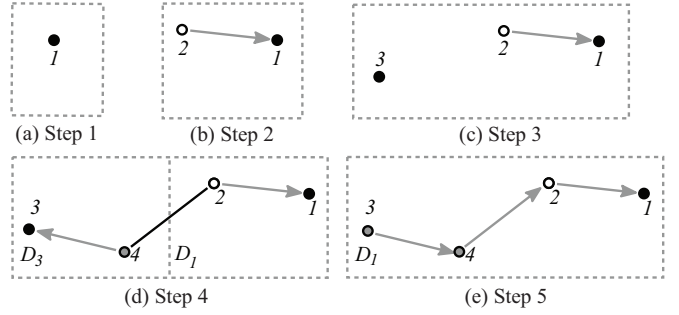} \\
    \caption{Starting the network.}
    \label{fig:bootstrap}
\end{figure}

In our architecture, starting the network is a bit more complex than that in a traditional WSN. Multiple BSs have to make a consensus on who should be the only active BS. In this subsection, we give a de-centralized solution to this problem.

Once a BS is booted, it is passive and listens for beacons from other sensor nodes. If, after some timeout, it still has not heard any beacon, it becomes active. Then, it connects to the server to retrieve UTC, and begins broadcasting beacons. Other sensor nodes receive the beacons and know their hop distances to the active BSs. Because sensor nodes use the gradient routing protocol, they join the nearest active BS. Notice that there might be several active BSs co-existing at this stage. The whole network is virtually split into several clusters, where each cluster has one active BS. 

Then, we discuss how the network merges these clusters in a de-centralized way. For simplicity of showing the merging process, we assume that the network only has two clusters $B_i$ and $B_j$ with the active BSs $b_i$ and $b_j$, respectively. There are obviously some nodes at the boundaries, belonging to one cluster and having neighbors belonging to the other one. These nodes can detect the presence of the two active BSs due to the beacon messages, as those belonging to $B_i$ will eventually hear about $b_j$ from their neighbors belonging to $B_j$. When these nodes detect the presence of the two active BSs, it is their duty to fix the problem. To keep things simple, we arbitrarily decide that the active BS with the smaller identifier should be kept active. Assuming $i < j$, the boundary nodes belonging to $B_j$ would thus send a \texttt{BS\_DOWN} message to $b_j$, asking it to become passive in favor of $b_i$.
Upon reception of this request, the BS $b_j$ stops sending beacons and becomes passive. As a result, routes to $b_j$ in the cluster $B_j$ gradually disappear, while at the same time routes to $b_i$ propagate from $B_i$ to $B_j$. When the process is over, the cluster $B_j$ has been merged with $B_i$, resulting in only one cluster. This merging process is also applicable when multiple clusters are present.


Figure~\ref{fig:bootstrap} provides an example of how the whole starting process operates. At the first step, BS $1$ is started. As it cannot hear from any other sensor node, it becomes active, gathering its own data and sending them to the server. Then, a regular sensor node $2$ is started. It detects the active BS $1$ and joins it to form a two-node network. At step three, BS $3$ is started. It is too far away to hear from BS $1$ and regular sensor node $2$, so it becomes active. At step four, another BS $4$ is added. It hears both from the active BS $3$ and from the regular sensor node $2$, and it decides to join BS $3$ rather than the small network $\{1, 2\}$ because of the shorter routing paths. Hence, there are two clusters: $B_1 = \{1, 2\}$ and $B_3 = \{3, 4\}$. The boundary nodes are regular sensor node $2$ and BS $4$, and respectively advertise about active BSs $1$ and $3$. When regular sensor node $2$ hears about BS $3$, it does nothing as regular sensor node $1$, its active BS, has a lower identifier than BS $3$. BS $4$, however, sends a \texttt{BS\_DOWN} message to BS $3$. Once BS $3$ becomes a passive BS and stops sending beacons, the route from BS $4$ to BS $3$ breaks, so that at some point, BS $4$ joins $B_1$, as well as BS $3$ later on, resulting in the final state of Figure~\ref{fig:bootstrap}.

\subsection{Gathering the information for adaptive selections}
\label{sec:gatherinfo}
{\changed Before adaptively selecting the active BSs, the network needs to learn the existence of other passive BSs and their battery levels. }For this purpose, we use a specific type of message, called \texttt{BS\_ADVERT}. The \texttt{BS\_ADVERT} messages are periodically generated by passive BSs, and then routed to the active BS like any other data message using gradient routing. The \texttt{BS\_ADVERT} messages are specifically handled. 
All sensor nodes include their own IDs in the packet when forwarding the \texttt{BS\_ADVERT} messages. When the active BS receives the  \texttt{BS\_ADVERT} messages, it knows exactly the paths that these messages have traveled through. By reverting these paths contained in the \texttt{BS\_ADVERT} messages, the active BS stores a \emph{handover table}, which is used when sending notifications to the next active BSs. This mechanism is well-known in ad hoc networks (\eg{} dynamic source routing~\cite{1994-WMCSA-JOHNSON}) and is sometimes called \emph{piggybacking}. 
The active BS also maintains a list of battery levels for all BSs. When the active BS receives a \texttt{BS\_ADVERT} message, it updates the corresponding elements in the list or table if this message contains newer timestamps. 

\begin{figure}
    \centering
    \includegraphics[width=\linewidth]{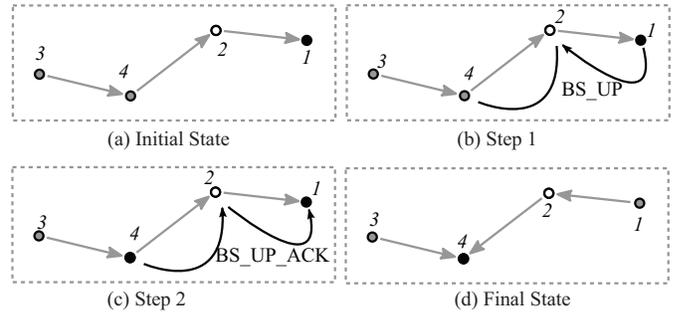} \\
    \caption{The handover process.}
    \label{fig:handover}
\end{figure}

\subsection{Handing over the active BS}
\label{sec:executedecision}
{\changed Knowing the locations of all passive BSs and their battery levels, the active BS will decide the next active BS based on the algorithm described in Section~\ref{sec:scheduling_algorithms}. }
If the active BS decides to hand over its role to another BS, it will send out a \texttt{BS\_UP} message for notifying its successor. This \texttt{BS\_UP} message contains the routing information from the handover table. 
Once a regular sensor node receives a \texttt{BS\_UP} message,
it forwards the message if it is on the route, and drops the message otherwise. 
When a BS receives the \texttt{BS\_UP} message, it checks whether it is the destination of the \texttt{BS\_UP} message. If it is, this BS sends back a \texttt{BS\_UP\_ACK} message to the currently active BS and becomes active by advertising its status through the beacon messages.
The previously active BS, upon reception of a \texttt{BS\_UP\_ACK}, becomes passive and stops its beacons. In the case where no  \texttt{BS\_UP\_ACK} is received (\eg{} node unreachable), the active BS tries again with the next best candidate. This process continues until a suitable candidate takes over the active role. 

The whole process of executing the handover decision is illustrated in Figure~\ref{fig:handover}. Initially, BS $1$ is active. It selects BS $4$ as its successor. During step $1$, a \texttt{BS\_UP} message is routed from BS $1$ to BS $4$ to inform BS $4$ the decision made by BS $1$. At step $2$, BS $4$ receives the \texttt{BS\_UP} message and becomes active. At the same time, it sends back a \texttt{BS\_UP\_ACK} message to BS $1$. Finally, BS $1$ becomes passive and BS $4$ is the only active BS.

{\changed \subsection{Recovering from failures}
\label{sec:recoverfromfaiulre}
}
{\changed In a sensor network, the active BS might fail and the network might split into small connected components. With our architecture, the network automatically recovers from these incidents.} When the active BS fails, it either reboots or stops working; both cases lead to the disappearance of active BS beacons. Should this happen, all routes in the network would disappear, and one or multiple BSs would eventually decide to become active, just like during the starting process. If multiples of them become active, the merging process would apply, eventually leading to only one active BS. {\changed When the network splits into small components, the passive BSs within each component are able to detect the disappearance of  beacons from the active BS in this component. Then, the  bootstrapping process mentioned in Section~\ref{sec:bootstrap-process} will ensure that there will eventually be one active BS in each small component.}

\section{Adaptive BS selection problem formulation}
\label{sec:formulation}
{\changed In the previous section, we have discussed the practical issues for coordinating multiple BSs.}  In this section, we consider the problem of optimally re-selecting the active BS, so that the energy resources on all BSs are efficiently utilized. {\changed We only consider the scenario where the network is fully connected. If the network splits into small components as we have seen in Section~\ref{sec:recoverfromfaiulre}, the problem is the same within each small component.}

Consider that $M$ BSs are deployed in the sensing field. Time is discretized into slots $n\in \mathbb{N}^+$, and we denote the length of a time slot by $\tau$. {\changed Notations are summarized in Table~\ref{tab:notation}.}

{\bf{Decision vector}}: 
As we have mentioned before, the active BS is adaptively re-selected in different time slots. Let ${v_m^{(n)}}$ indicate whether BS $m$ is active during a given time slot $n$, \textit{i.e.},
${v_m^{(n)}}=\mathbb{I}\left(\mbox{BS $m$ is active during time slot $n$}\right)$,
where $\mathbb{I}(A)$ denotes the indicator function: $\mathbb{I}(A)=1$ if argument $A$ is true and $\mathbb{I}(A)=0$ otherwise.
Collect all ${v_m^{(n)}}$, $1\leq m\leq M$, in an $M \times 1$ column vector
$\vectw{v}{(n)}=\left[v_1^{(n)} \; v_2^{(n)} \; \cdots \; v_M^{(n)}\right]^\top$ with $\top$ denoting transposition.
{Call $\vectw{v}{(n)}$ the \emph{decision vector} during time slot $n$. 
Because only one active BS is possible during one time slot, $\vectw{v}{(n)}$ has $M-1$ zero entries and one entry equal to $1$. } 
We denote the sequence of decision vectors up to time slot $n$ by $\mathcal{V}^{(n)}=\left\{\bm{v}^{(t)}\right\}_{t=1}^n$.


{\bf{Cost matrix}}:
The energy consumption of BSs might come from three parts: sensing, short-range communication, and long-range communication. We assume that the sensing costs are negligible. Let the MAC protocol and routing protocol of the WSN be predefined. Therefore, when a specific BS is selected to be active, both the energy consumption from short-range communication and from long-range communication of each BS is deterministic.\footnotemark \footnotetext{In practice, the energy consumption rates might have slight deviations given that the active BS is selected.} Denote by $C_{ml}$ the energy consumption rate of BS $m$ $(1\leq m\leq M)$ when BS $l$ $(1\leq l\leq M)$ is active. We group all energy consumption rates in an $M \times M$ matrix $\bm{C}$, which we call the \emph{cost matrix}. If we neglect the energy consumption from short-range communication, the passive BSs do not consume any energy, and therefore the cost matrix becomes diagonal. In practice, the ratio between the energy consumption from long-range communication and that from short-range communication might be $5\sim 20$, based on different settings of the network.

{\bf{Available energy}}: We denote the remaining  amount of energy of BS $m$ at the end of time slot $n$ by $e_m^{(n)}$ and we call it \emph{available energy}. We gather the available energy of all BSs in a vector 
$\vectw{e}{(n)}=\left[e_1^{(n)} \; e_2^{(n)} \; \cdots \; e_M^{(n)}\right]^\top
$.
In practice, available energy is lower-bounded by zero and upper-bounded by the storage capacity. In the analysis of this paper, however, we assume that it is not upper-bounded for simplicity. Without loss of generality,  we assume that all BSs have the same available energy {\changed $e_0$} initially, with {\changed  $\vectw{e}{(0)}=e_0 \bm{u}_M$}, where $\bm{u}_M=\left[1\, 1\, \cdots \,1\right]^\top$ is the $M\times 1$ all-ones vector.

{
\begin{table}[t]
\renewcommand{\arraystretch}{1.3}
\begin{center}
    \caption{Notations}
  \begin{tabular}{ | p{0.2\columnwidth} | p{0.65\columnwidth}| }
     \hline
     $\tau$ & Length of a time slot\\
     $M$ & Number of BSs\\
               $N$ & Lifetime of the network\\
          $\bm{u}_M$ & The uniform vector with $\bm{u}_M=[1,1,\cdots 1]^\top$\\
     $\bm{v}^{(n)}$ & Decision vector during time slot $n$\\
               $\bm{C}$ & Cost matrix\\
     $\bm{e}^{(n)}$ & Available energy during time slot $n$\\
          $e_0$ & Initial available energy of all BSs\\
          $\bm{s}^{(n)}$ & Energy recharge rates during time slot $n$\\
                    $\bm{\bar{s}}$ &  Energy consumption rates in expectation\\
    \hline
  \end{tabular}
      \label{tab:notation}
    \end{center}
\end{table}
}

{\bf{Energy recharge rates}}: 
During each time slot $n\in \mathbb{N}^+$, each BS $m$ ($1\leq m \leq M$) receives a certain amount of incoming energy. Denote the average rate of incoming energy during this time slot by $s_m^{(n)}$ and call it the \emph{energy recharge rate}. We group all the energy recharge rates during time slot $n$ into a vector
$
\vectw{s}{(n)}=\left[s_1^{(n)} \; s_2^{(n)} \; \cdots \; s_M^{(n)}\right]^\top.
$
We denote the sequence of energy recharge rates up to time slot $n$ by
$\mathcal{S}^{(n)}=\left\{\bm{s}^{(t)}\right\}_{t=1}^n$.
In particular, $\mathcal{S}^{(\infty)}$ denotes the sequence of energy recharge rates over an infinite time horizon. 
We make the following realistic assumptions on  $\mathcal{S}^{(\infty)}$:
\begin{itemize}
\item D1:
\begin{equation}
\mathbb{E}_{n-1} \bm{s}^{(n)}= \bar{\bm{s}}, \forall n\in \mathbb{N}^+,
\label{eq:recharge_rate_constant_expectation}
\end{equation}
where $\bar{\bm{s}}$ is a constant vector {\changed  and $\mathbb{E}_{n-1}$ is a shorthand for denoting the expectation conditioned on the sequence $\mathcal{S}^{(n-1)}$, that is, $\mathbb{E}_{n-1} \bm{s}^{(n)}= \mathbb{E} (\bm{s}^{(n)} | \mathcal{S}^{(n-1)} )$. 
 Let $\bar{s}_m$ be the $m$-th element of ${\bm{\bar{s}}}$}.
This assumption is weaker than assuming  $\mathcal{S}^{(\infty)}$ is an i.i.d process. 
\item D2:
\begin{equation}
 \left\|\bm{s}^{(n)}\right\|_{\infty} \leq S, \forall n\in \mathbb{N}^+.
\label{eq:recharge_rate_boundedness}
\end{equation}
where $S$ is a constant with $0 \leq S < +\infty$. 

\end{itemize}

{\bf{Relations among the aforementioned parameters}}:
During time slot $n$, the amounts of energy recharged for all BSs given by $\tau \bm{s}^{(n)}$ and the amounts of energy consumed are given by $\tau \bm{C}\cdot \bm{v}^{(n)}$. Therefore, the available energy evolves according to
\begin{equation}
\vectw{e}{(n)}=\vectw{e}{(n-1)}+\tau  \bm{s}^{(n)}-\tau \bm{C}\cdot \bm{v}^{(n)}.
\label{eq:iterative_function}
\end{equation}
{\changed  If we sum up the iterative equation (\ref{eq:iterative_function}) from time $0$ to time $n$ and use $\bm{e}^{(0)}= e_0 \bm{u}_M$, we have
\begin{equation}
\bm{e}^{(n)}=e_0 \bm{u}_M+  \tau \sum_{t=1}^{n} \bm{s}^{(t)} - \tau\bm{C}\cdot \sum_{t=1}^{n} \bm{v}^{(t)}.
\label{eq:sum_iterative_function}
\end{equation}
}

{\bf{Adaptive BS selection problem:}} 
{
\changed 
Denote the lifetime of the network by $N$.
If the realization of $\mathcal{S}^{(\infty)}$ is already known to us, the goal is to schedule the selections of active BSs, such that the lifetime $N$ is maximized. In other words, we want to find the longest sequence of decision vectors $\mathcal{V}^{(N)}$ such that for any $1\leq n \leq N$, the available energy $\bm{e}^{(n)} \geq \bm{0}$.\footnotemark \footnotetext{
Without special mentioning in this paper, the inequalities between vectors are all component-wise.} Therefore, we  
formulate the problem as an optimization problem
\begin{align}
\nonumber &\max_{\mathcal{V}^{(N)} } && N\\
\nonumber & \text{s.t.} &&\tau\bm{C}\cdot \sum_{t=1}^{n} \bm{v}^{(t)} \leq e_0 \bm{u}_M+\tau \sum_{t=1}^{n} \bm{s}^{(t)} , \forall 1\leq n \leq N,\\
\label{eq:max_lifetime} 
&&& \bm{u}_M^\top\cdot {\bm{v}}^{(n)}=1, \ \forall 1\leq n \leq N,\\
\nonumber &&& \bm{v}^{(n)} \in \{0,1\}^M, \ \forall 1\leq n \leq N,
\end{align}
where the first constraint follows from (\ref{eq:sum_iterative_function}) and that $\bm{e}^{(n)} \geq \bm{0}, \forall 1\leq n\leq N$.

We denote the optimal objective value of problem  (\ref{eq:max_lifetime}) by $N_{\rm{opt}}$. We denote the offline scheduling algorithm that optimizes (\ref{eq:max_lifetime}) by \emph{OPT}. We will use it for comparisons in Section~\ref{sec:simulations}.
We note that: (i) problem (\ref{eq:max_lifetime}) is not a standard optimization problem because the number of constraints is dependent on the objective value $N$. (ii) The optimal lifetime $N_{\rm{opt}}$ depends on the realization of the stochastic process $\mathcal{S}^{(\infty)}$. In the following, we will analyze the asymptotic performance of the optimal objective value $N_{\rm{opt}}$ when $e_0\rightarrow \infty$ through an auxiliary optimization problem.

Denote the fraction of active time of BS $m$ ($1\leq m\leq M$) by $\bar{v}_m$. Group these fractions into a vector
$\bar{\bm{v}}=\left[ \bar{v}_1\; \bar{v}_2\; \cdots \; \bar{v}_M\;\right]^\top$. Notice that we have $\bm{u}_M^\top \bm{\bar{v}}=1$. We denote by  
\begin{equation}
\bm{R}= \bm{C}-\bar{\bm{s}}\cdot \bm{u}_M^\top.
\label{eq:defR}
\end{equation}
Given the fractions of active time $\bm{\bar{v}}$, the expected energy decrease rates of all BSs are $\bm{C}\bm{\bar{v}}-\bar{\bm{s}}$, which are equivalent to $\bm{R}\bm{\bar{v}}$ because of (\ref{eq:defR}) and $\bm{u}_M^\top \bm{\bar{v}}=1$. Because the lifetime of the network is decided by the maximum energy decrease rate among all BSs, maximizing the lifetime amounts to minimizing the maximum energy decrease rate. Therefore, to analyze the asymptotic property of the optimal lifetime $N_{\rm{opt}}$, we define the auxiliary optimization problem 
\begin{equation}
\begin{aligned}
&\min_{\bar{\bm{v}},f}
& & f\\
& \text{s.t.} & & \bm{R}  \bar{\bm{v}} \leq f \bm{u}_M,\\
&&& \bm{u}_M^\top \bar{\bm{v}}=1,\\
&&&\bar{\bm{v}}\geq \bm{0},
\end{aligned}
\label{eq:max_dropping_rate}
\end{equation}
whose optimal solution is denoted by  $(\bar{\bm{v}}^*, f^*)$. 

In the following, we will show the relation between the optimal objective value $N_{\rm{opt}}$ of problem (\ref{eq:max_lifetime}) and  the optimal objective value $f^*$ of problem (\ref{eq:max_dropping_rate}) under assumptions D$1$ and D$2$: (i) If $f^*<0$, by selecting the active BSs according to the optimal fractions $\bar{\bm{v}}^*$, the available energy of all BSs has a tendency to increase with time. Therefore, the probability that the optimal lifetime $N_{\rm{hef}}$ is infinite converges to $1$ when $e_0\rightarrow \infty$. (ii) If $f^*>0$, any scheduling algorithm will result in a finite lifetime almost surely. By selecting the active BSs according to the optimal fractions $\bar{\bm{v}}^*$, there is a high probability that the optimal lifetime is within the range $[(1-\delta)e_0/\tau f^*, (1+\delta)e_0/\tau f^*]$, for any $\delta>0$. This probability converges to $1$ when $e_0\rightarrow \infty$. The arguments above are summarized in Theorem~\ref{lm:relation_between_opts}.

\begin{theorem}
If assumptions D$1$ and D$2$  on the energy recharge rates $\mathcal{S}^{(\infty)}$ are met, the optimal objective value 
$N_{\rm{opt}}$ of problem (\ref{eq:max_lifetime}) has the following asymptotic performance:
\begin{itemize}
\item when $f^*< 0$, 
\begin{equation}
\lim_{e_0 \rightarrow \infty} \mathbb{P}  \left(  N_{\rm{opt}} =\infty\right)=1,
\label{eq:thm1r2}
\end{equation}

\item when $f^*>0$, 
\begin{equation}
\lim_{e_0 \rightarrow \infty} \mathbb{P} \left( \left|  \frac{N_{\rm{opt}}}{e_0/(\tau f^*)}-1\right|<\delta \right)=1, \forall \delta>0.
\label{eq:thm1r1}
\end{equation}
\end{itemize}
\label{lm:relation_between_opts}
\end{theorem}

The detailed proof is found in the technical report~\cite{1403.2527}, which we briefly sketch here. In the simple deterministic scenario where the energy recharge rates $\bm{s}^{(n)}=\bar{\bm{s}}$ for any $n \in \mathbb{N}$, we can easily show that:
given that $f^*<0$, if $e_0$ is sufficiently large, $N_{\rm{opt}}=\infty$; and given that $f^*>0$, if $e_0$ is sufficiently large, $N_{\rm{opt}}$ is deterministically within the range $[(1-\delta)e_0/\tau f^*, (1+\delta)e_0/\tau f^*]$, for any $\delta>0$.
Notice that the total amount of energy recharged in the first $n$ time slots in the deterministic scenario is $n \bm{\bar{s}}$ and that in the stochastic scenario is $\sum_{t=1}^n \bm{s}^{(t)}$. Under assumptions D$1$ and D$2$, we show that their difference $\sum_{t=1}^n \bm{s}^{(t)}- \bm{\bar{s}}$ is a martingale with bounded difference.
We use the Azuma-Hoeffding inequality for martingales~\cite[p. 476]{Geoffrey01} to show that the probability distribution of the distance from $\sum_{t=1}^n \bm{s}^{(t)}- \bm{\bar{s}}$ to the zero vector decays exponentially. Using this result, we will show that when $e_0 \rightarrow \infty$, the optimal lifetime $N_{\rm{opt}}$ in the stochastic scenario converges in probability to that in the simple deterministic scenario.
}

Solving (\ref{eq:max_lifetime}) or (\ref{eq:max_dropping_rate}) is however infeasible in practice because of the following reasons: (i) measuring the cost matrix $\bm{C}$ requires expensive equipments such as high-frequency data loggers and (ii) estimating the energy recharge rates  $\mathcal{S}^{(n)}$ is hard, because they depend on too many factors. For example, the energy recharge rate from a solar panel might depend on its location, the angle of its surface to the sunlight, its energy conversion efficiency, and the weather.
In a real WSN, the only easy-to-capture information is the battery level, which can be used as an indicator of the available energy. In the following, we will discuss an algorithm for re-selecting the active BS which only uses information on available energy as input.

\section{The ``Highest Energy First'' (HEF) Algorithm}
\label{sec:scheduling_algorithms}
In this section, we propose the algorithm \emph{``Highest Energy First''} (HEF) for solving the adaptive BS selection problem. {\changed In practice, this algorithm  is easy to implement because it only requires the battery levels of all BSs as the input.} 

The procedure of running HEF is summarized in Algorithm~\ref{alg:hef}.
At any time slot $n$, BS $m^*$ ($1\leq m^*\leq M$) is chosen to be active during time slot $n$ if and only if its available energy $e_{m^*}^{(n-1)}$ is the highest, \textit{i.e.},
\begin{equation}
\begin{aligned}
v_{m^*}^{(n)}&=\mathbb{I}{\left( e_{m^*}^{(n-1)}  \geq e_m^{(n-1)}, \ \forall 1\leq m \neq {m^*} \leq M\right)},
\end{aligned}
\label{eq:def_hef}
\end{equation}
with ties broken uniformly at random.

{
\changed 
Let $N_{\rm{hef}}$ be the lifetime of the network using the HEF scheduling algorithm, that is, 
\[
N_{\rm{hef}}=\inf \{\{\infty\}\cup  \{n \mid \exists 1\leq l^*\leq M,  e_{l^*}^{(n+1)}< 0\}\}.
\]

The HEF algorithm is a heuristic algorithm, yet we will show that it is asymptotically optimal under mild conditions. We use the optimal objective value $f^*$ of problem (\ref{eq:max_dropping_rate}) as a link between $N_{\rm{hef}}$ and $N_{\rm{opt}}$: 
 (i) If $f^*<0$, for any large constant $K$, there is a high probability that the lifetime $N_{\rm{hef}}>Ke_0$ when the initial available energy $e_0$ is large. This probability converges to $1$ when $e_0\rightarrow \infty$. This result is a bit weaker than that $\lim_{e_0 \rightarrow \infty} \mathbb{P} (N_{\rm{hef}}=\infty)=1$ as in (\ref{eq:thm1r2}). (ii) If $f^*>0$, when $e_0$ is large, there is a high probability that $N_{\rm{hef}}$ is within the range $[(1-\delta)e_0/\tau f^*, (1+\delta)e_0/\tau f^*]$, for any $\delta>0$. This probability converges to $1$ when $e_0\rightarrow \infty$. We summarize the arguments above in Theorem~\ref{thm:hef_opt}.
 
\begin{theorem}
If assumptions D$1$ and D$2$  on the energy recharge rates $\mathcal{S}^{(\infty)}$ are met, and if in addition
\begin{itemize}
\item{D3:} $R_{ij}=C_{ij}-\bar{s}_i<0$, $\forall 1\leq i \neq j \leq M$, and
\item{D4:} $(\bm{C}^{\top})^{-1}\bm{u}_M>\bm{0}$,
\end{itemize}
then 
\begin{itemize}
\item when $f^*< 0$, 
\begin{equation}
\forall K, \lim_{e_0 \rightarrow \infty} \mathbb{P}  \left(  N_{\rm{hef}} > K e_0\right)=1,
\label{eq:thmresult2}
\end{equation}
\item when $f^*>0$, 
\begin{equation}
\lim_{e_0 \rightarrow \infty} \mathbb{P} \left( \left|  \frac{N_{\rm{hef}}}{e_0/(\tau f^*)}-1\right|<\delta \right)=1, \forall \delta>0.
\label{eq:thmresult1}
\end{equation}
\end{itemize}
\label{thm:hef_opt}
\end{theorem}
}

\begin{algorithm}[t]
\caption{The ``Highest Energy First'' Algorithm}
\label{alg:hef}
\begin{algorithmic}
\Require $\bm{e}^{(0)}, \mathcal{S}^{(n)}$
\Ensure $\mathcal{V}^{(n)}$
\For{$t=1\ \rm{  to  }\ n$}
\State Find $m^*$ such that $e_{m^*}^{(t-1)} \geq e_m^{(t-1)}$, $\forall 1\leq m \neq m^* \leq M$
\State Set $\bm{v}^{(t)}$ where $v_{m^*}^{(t)} \gets 1$ and $v_m^{(t)} \gets 0$, for any $m \neq m^*$.
\State Update $\vectw{e}{(t)}=\vectw{e}{(t-1)}-\tau \bm{C}\cdot \bm{v}^{(t)}+\tau  \bm{s}^{(t)}$.
\EndFor
\end{algorithmic}
\end{algorithm}
{
\changed 
We interpret conditions D$3$ and D$4$ in Theorem~\ref{thm:hef_opt} as follows: (i) Condition D$3$ states that for any passive BS, the expected energy recharge rate is larger than the energy consumption rate, regardless of the selection of the active BS. 
(ii) Condition D$4$ is satisfied  when energy consumption rates of active BSs (diagonal elements of $\bm{C}$) are much larger than the differences among the energy consumption rates of all passive BSs (differences among non-diagonal elements of $\bm{C}$).
Indeed, we define  $c_{\rm{pb}}= \min_{1\leq i \neq j\leq M} C_{ij}$ and decompose $\bm{C}$ as $\bm{C}= \bm{\Lambda}+c_{\rm{pb}} \bm{u}_M \bm{u}_M^\top$. Then, the diagonal elements of $\bm{\Lambda}$ are much larger than the non-diagonal elements. It follows that $\bm{\Lambda}$ is near diagonal and therefore  $(\bm{\Lambda}^\top)^{-1}\bm{u}_M>\bm{0}$. Using the Sherman-Woodbury-Morrison identity\footnotemark \footnotetext{\changed The Sherman-Woodbury-Morrison identity states that for any matrix $\bm{A}$ and for any two vectors $\bm{w}_1$ and $\bm{w}_2$, if $1+\bm{w}_2^\top \bm{A}^{-1} \bm{w}_1\neq 0$, we have $(\bm{A}+\bm{w}_1 \bm{w}_2^\top)^{-1}= \bm{A}^{-1}- (\bm{A}^{-1} \bm{w}_1 \bm{w}_2^\top \bm{A}^{-1})/(1+\bm{w}_2^\top \bm{A}^{-1}\bm{w}_1)$.}, we see that
\[
(\bm{C}^\top)^{-1}\bm{u}_M= (\bm{\Lambda}^\top)^{-1}\bm{u}_M/(1+c_{\rm{pb}} \bm{u}_M^\top \bm{\Lambda}^{-1}\bm{u}_M )>\bm{0}.
\]
More justifications of conditions D$3$ and D$4$ through simulations are shown in Section~\ref{subsec:validation_opt_condition}.

The detailed proof is found in the technical report~\cite{1403.2527}. Here, we sketch the intuition for the proof: (i) First, we show that with condition D$3$, there is a high probability that all BSs use up their available energy at time $N_{\rm{hef}}+1$ when $e_0$ is large. (ii) Secondly, we show that the event that all BSs use up the energy at time $N_{\rm{hef}}+1$  implies that the average decision vector $\sum_{n=1}^{N_{\rm{hef}}+1} \bm{v}^{(n)}/(N_{\rm{hef}}+1)$ converges to ${\bm{R}^{-1} \bm{u}_M  }/{\bm{u}_M^{\top}  \bm{R}^{-1} \bm{u}_M }$ in probability. Under condition D$4$, we show that the optimal solution of problem (\ref{eq:max_dropping_rate}) is $\bm{\bar{v}}^*={\bm{R}^{-1} \bm{u}_M  }/{\bm{u}_M^{\top}  \bm{R}^{-1} \bm{u}_M }$. 
(iii) Thirdly, given that  the average decision vector converges in probability to the optimal solution $\bm{\bar{v}}^*$ of problem (\ref{eq:max_dropping_rate}), we use the Azuma-Hoeffding inequality and deduce that: if $f^*<0$, there is a high probability that $N_{\rm{hef}}>Ke_0$; and if $f^*>0$, there is a high probability that $N_{\rm{hef}}>(1-\delta)e_0/(\tau f^*)$. Noticing that $N_{\rm{hef}} \leq N_{\rm{opt}}$ and (\ref{eq:thm1r2}), we conclude the proof.

}

\section{Simulations}
\label{sec:simulations}
In this section, we will show how we evaluate the proposed scheme by running several simulations\protect \footnotemark~on the simulator Castalia/OMNeT++ \cite{4351800}. 

\begin{table}[t]
\renewcommand{\arraystretch}{1.3}
\begin{center}
    \caption{Simulation settings}
  \begin{tabular}{ | p{0.45\columnwidth} | p{0.45\columnwidth}| }
     \hline
     Sensing field & $\changed 200\,\rm{m} \times 200\, \rm{m}$\\
     Sensor node positions & uniformly at random\\
     Radio layer model& $\rm{XE}1205$ chip, unit disk model, the transmitting range is $40\, \rm{m}$\\
     Radio energy consumptions in TX$\backslash$RX$\backslash$Sleep mode& $79.45\backslash 46 \backslash 1.4\,\rm{mW}$\\
      Data generating rate & $1\,\rm{packet}/\rm{sec}$\\
     Control message rate & $1\,\rm{packet}/5\, \rm{min}$\\
     GSM/GPRS connection rate & once/$5\, \rm{min}$\\
     Average power consumption of GSM/GPRS per connection& $296\,\rm{mW} \times 40\,\rm{sec}$ \\
     Active BS handover interval & every $2\,\rm{hours}$\\
     Initial available energy & $14400\,\rm{J}$ \\
     Solar panel & $50\,\rm{cm}^2$\\
    \hline
  \end{tabular}
      \label{tab:simulation_parameters}
    \end{center}
\end{table}

\subsection{General Settings}
The general settings of the simulations are chosen to closely approximate our hardware specifications, as listed in Table~\ref{tab:simulation_parameters}.
{\changed We simulate a sensor network composed of $5$ BSs ($M=5$) and $35$ regular sensor nodes, which are distributed uniformly at random in a $200 \rm{m}\times 200 \rm{m}$ sensing field. }
In the physical layer of all sensor nodes, we simulate the $\rm{XE}1205$ radio transceiver, with the transmitting power fixed to $0\,\rm{dbm}$. We adopt the ideal unit disk model for the wireless channel and choose the parameters so that the transmitting range is fixed to $40\, \rm{m}$. In the MAC layer, the T-MAC protocol is used.
All sensors generate data packets at a rate of $1\, \rm{packet/sec}$. The \texttt{BS\_ADVERT} message (Section~\ref{sec:architecture}) is transmitted at a rate of $1\, \rm{packet}/5\, \rm{min}$. {\changed Then, the energy consumption rates of sensor nodes for using the short-range communications are captured using the built-in modules of the simulator Castalia/OMNeT++.}
The active BS connects to the remote server with GSM/GPRS every $5\, \rm{min}$. Because the transmitted data volume during each connection is small, the major part of the energy consumption comes from starting, maintaining and closing the communication. We assume that for each GSM/GPRS connection, the active time and the average power consumption is $40\,\rm{sec}$ and $296\,\rm{mW}$ (we choose these values based on the measurements with a digital oscilloscope). The active BS decides whether to transfer its role every $2$ hours, which amounts to $\tau= 2\, \mbox{hours}$ for each time slot. 
Each BS is assumed to have a set of AA NiMH rechargeable batteries with an initial energy of $800\, \rm{mAh}\times 5\,\rm{V}=14400\,\rm{J}$ and a solar panel. 
{\changed 
We assume that the energy recharge rate for BS $m$ ($1\leq m\leq M$) during time slot $n\in \mathbb{N}$ is 
\[
s_m^{(n)}= \eta_m \cdot \gamma_m \cdot I_m^{(n)} \cdot \Gamma_{\rm{default}},
\]
where $\eta_m$ denotes the energy conversion efficiency of the solar panel for BS $m$, $\gamma_m$ denotes the coefficient for losses (inverter loss, temperature loss, energy transmission loss, energy conservation loss and low radiation loss), $I_m^{(n)}$ denotes the solar radiation on BS $m$ at time $n$, and $\Gamma_{\rm{default}}$ is the default size of the solar panel.
The solar radiations $ \{I_m^{(n)}\}_{n}$ ($1\leq m \leq M$) we use are captured from project Sensorscope~\cite{2010-TOSN-INGELREST}. We set $\Gamma_{\rm{default}}=50\,\rm{cm}^2$. For all $1\leq m \leq M$, we let $\eta_m$ be drawn from $[0.05,0.15]$ uniformly at random, and we set $\gamma_m=0.2$.
}
The settings discussed above are default unless other settings are explicitly mentioned.

\begin{figure}[t]
    \centering
        \includegraphics[width=0.98\linewidth]{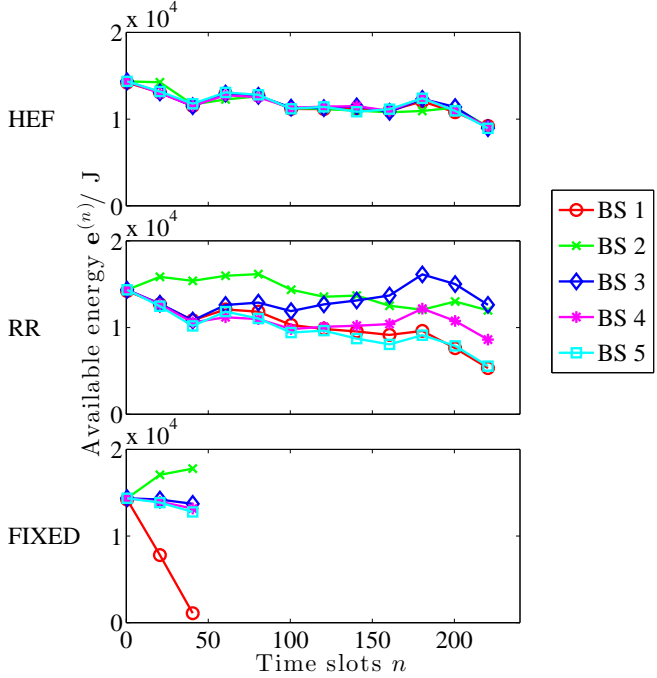}
    \caption{The available energy $\bm{e}^{(n)}$ ($1\leq n \leq 240$) when running different algorithms for selecting active BSs. FIXED depletes the battery of the only active BS quickly, thus leading to an early death of the WSN. RR cannot fully utilize all the energy because different BSs can have very different energy recharged from solar panels. HEF equalizes the available energy of all BSs despite different energy harvested on different BSs. It can substantially prolong the lifetime of the WSN compared to FIXED and {RR}.}
    \label{fig:energy_efficiency}
\end{figure}

\subsection{Performance of Different Algorithms}
\label{subsec:performance}
In the following, we show the performances of four different algorithms for organizing the WSN, \textit{i.e.}, FIXED, {\changed \emph{Round-Robin} (RR)}, OPT and HEF. FIXED denotes the scheme with the active BS fixed to be BS $1$.  {\changed RR} denotes the algorithm where all BSs take turns to be active and have perfectly identical active times. {\changed OPT is the offline optimal scheduling algorithm. It is not applicable in practice and is only used for comparison. From the simulator Castalia/Omnet++, we get the energy consumption rates of all sensor nodes when different BSs are active. We list these energy consumption rates into the cost matrix $\bm{C}$. Then, we solve the optimization problem (4) and have the optimal selections of active BSs.}
Finally, HEF is the ``highest energy first'' algorithm described in Section~\ref{sec:scheduling_algorithms}.
In the following, we will compare their performances in different aspects. {\changed To avoid the simulation to run infinitely long time, we restrict the maximum running time to be $2400$ time slots ($200$ days): if a network can sustain $2400$ time slots, we consider its lifetime as infinite.}


{\changed \bf{Available energy versus time}}: First, we show the available energy $\bm{e}^{(n)}$ {\changed during $20$ days ($1\leq n\leq 240$)},  when running different algorithms in Figure~\ref{fig:energy_efficiency}. We see that HEF leads $\bm{e}^{(n)}$ to be uniform despite different energy harvested for different BSs. RR cannot fully utilize all the energy because different BSs can have very different energy recharged from solar panels. FIXED leads to a fast energy decrease rate of the only active BS, resulting in an early death of the WSN.

\begin{figure}[t]
    \centering
    \includegraphics[width=\linewidth]{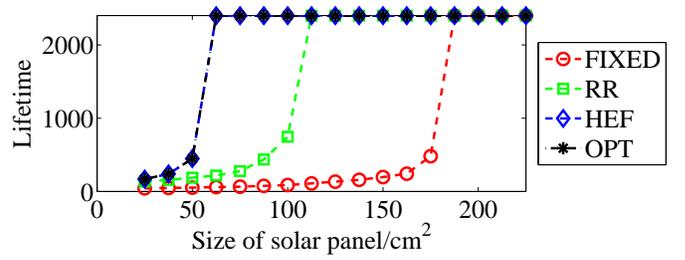}
    \caption{\changed Lifetime versus size of the solar panels. The minimum sizes of solar panels to achieve an infinite lifetime in a network running HEF, RR and FIXED are $62.5\rm{cm}^2$, $112.5\rm{cm}^2$ and $187.5\rm{cm}^2$, respectively.}
    \label{fig:LifetimeSolarSize}
\end{figure}

{\changed {\bf{Lifetime versus size of solar panels}}: In Figure~\ref{fig:LifetimeSolarSize}, we show that the lifetime of the network increases with the size of the solar panel equipped on each sensor node. When the size of the solar panel is large enough, the lifetime becomes infinite.
The minimum sizes of solar panels to achieve an infinite lifetime in a network running HEF, RR and FIXED are $62.5\rm{cm}^2$, $112.5\rm{cm}^2$ and $187.5\rm{cm}^2$, respectively. The lifetime of HEF is always better than that of RR and FIXED, and is close to that of OPT.
}

\begin{figure}[t]
    \centering
    \includegraphics[width=\linewidth]{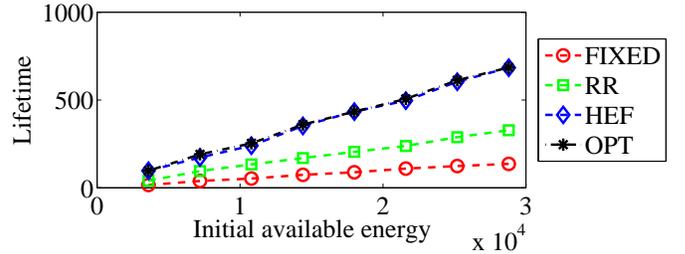}
    \caption{\changed Lifetime versus initial available energy. We see that the lifetime of HEF and OPT increases linearly with the amount of the initial available energy. HEF is always close to OPT and is better than RR and FIXED.}
    \label{fig:LifetimeBatteryCap}
\end{figure}

{\changed {\bf{Lifetime versus initial available energy}}: In Figure~\ref{fig:LifetimeBatteryCap}, we show how the lifetime changes when the sensor network is given different amount of initial available energy $e_0$. Here we set the size of all solar panels to be $50\rm{cm}^2$, which is not sufficient for the network to have an infinite lifetime when running any algorithm. We see that in this scenario, the lifetime of both OPT and HEF increases linearly with the initial available energy, as indicated by the arguments used to prove Theorem~\ref{lm:relation_between_opts} and~\ref{thm:hef_opt} when $f^*>0$. HEF is close to OPT and is better than RR and FIXED. 
}

\begin{figure}[t]
    \centering
    \includegraphics[width=\linewidth]{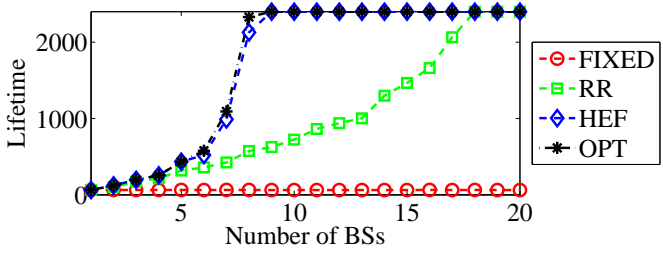}
    \caption{\changed  Lifetime versus number of BSs. We see that the lifetime when running HEF or RR increases with the number of BSs. The number of BSs to sustain an infinite lifetime required by HEF and RR are $9$ and $18$, respectively. When running FIXED, larger number of BSs does not result in longer lifetime because the burden is not shared among all BSs.}
    \label{fig:LifetimeBSNum}
\end{figure}

{\changed {\bf{Lifetime versus number of BSs}}: In Figure~\ref{fig:LifetimeBSNum}, we show how the lifetime changes when the sensor network has different number of BSs $M$. 
We see that when running HEF or RR, the lifetime increases with the number of BSs. This is because a large number of installed BSs will average out the high cost of being the active BS. On the contrary, the lifetime of FIXED remains constant when the number of BSs increases because the burden of using long-range communication is not shared among all BSs. From Figure~\ref{fig:LifetimeBSNum}, we see that the number of BSs to sustain an infinite lifetime required by HEF and RR are $9$ and $18$, respectively.
}

\begin{figure}[t]
    \centering
    \includegraphics[width=\linewidth]{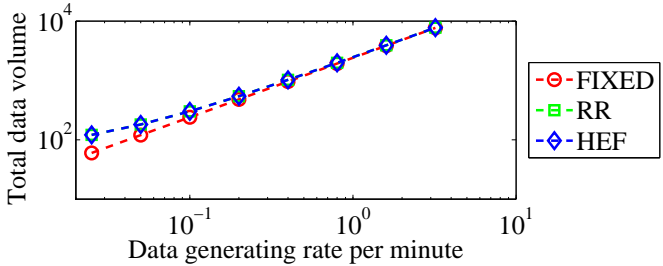}
    \caption{Overall number of packets transmitted per hour versus the sensing rate of each sensor node. We see that the communication overheads of both {\changed RR} and HEF are very small.}
    \label{fig:communication_overhead}
\end{figure}

\begin{figure*}[ht]
    \centering
     \subfloat[]{\label{figsub:adap1}\includegraphics[width=.49\linewidth]{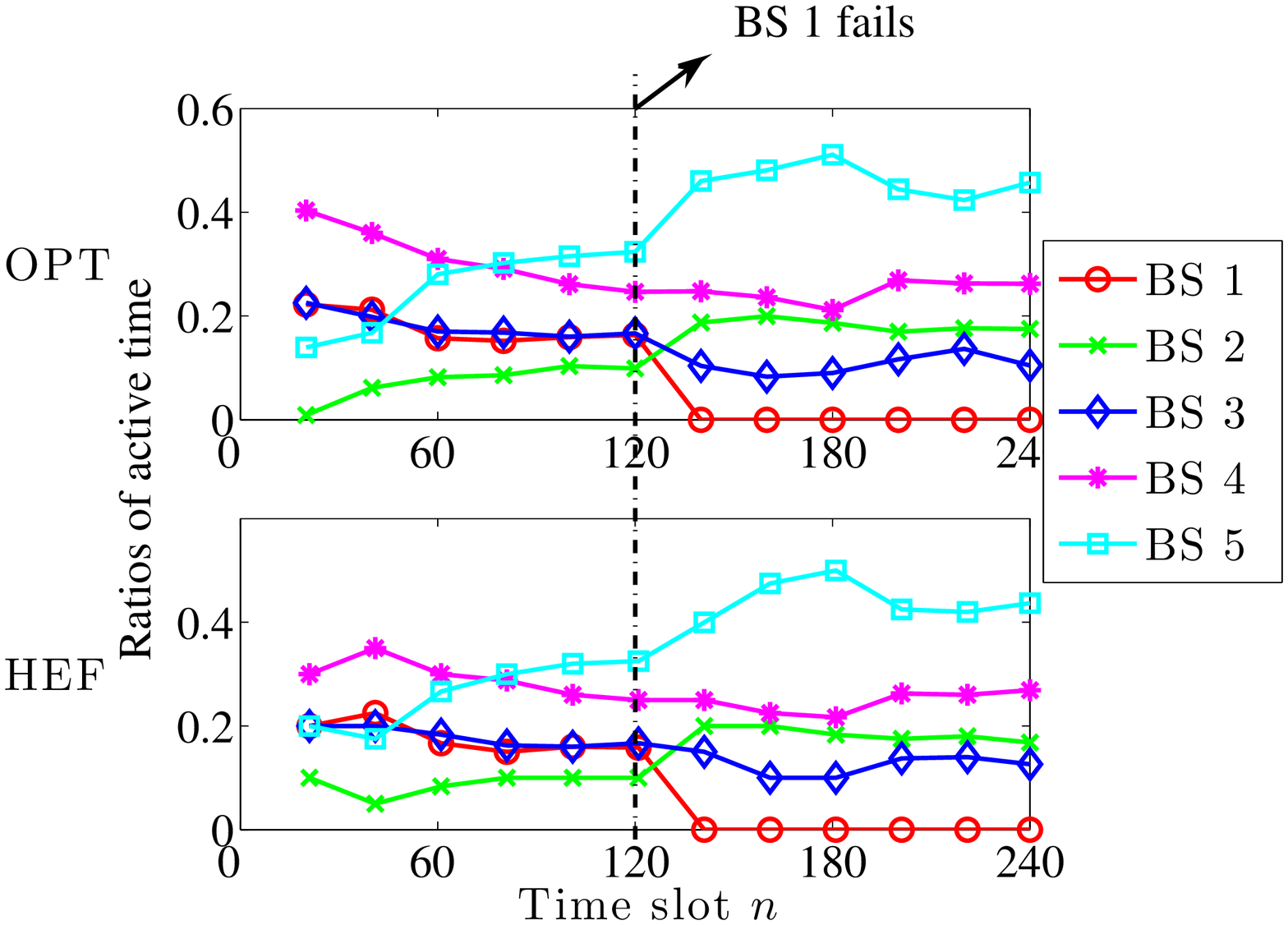}}\hfil
        \subfloat[]{\label{figsub:adap2}       \includegraphics[width=.49\linewidth]{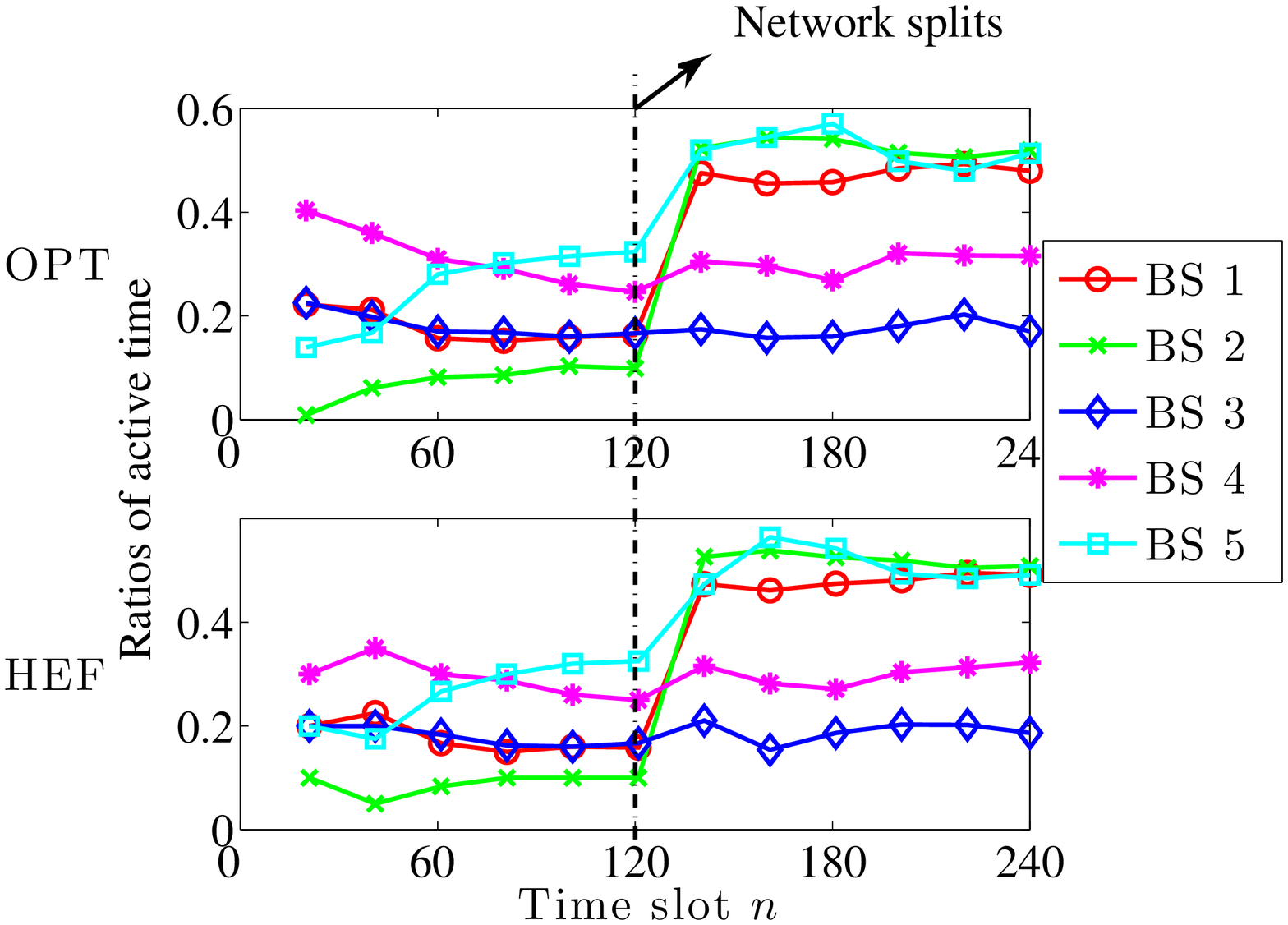}}\hfil
    \caption{\changed The reactions to network changes when running HEF. We use the ratios of active time for all BSs as a metric. Figure~\ref{figsub:adap1} shows the scenario where BS $1$ fails at time slot $n=120$. Figure~\ref{figsub:adap2} shows the scenario where the network splits into two small components (one component has BS $1$ and BS $2$ and the other component has BS $3$, BS $4$ and BS $5$) at time slot $n=120$. We see that in both scenarios, HEF reacts rapidly to network changes and always closely follows OPT.}
    \label{fig:adaptability_annotated}
\end{figure*}

{\changed To sum up, HEF is more energy-efficient than RR and FIXED, and it is very close to OPT in all simulated scenarios. We list the results in the second column of Table~\ref{tab:comparision_different_scheduling_algorithms}.
}

{\changed  {\bf{Reactions to network changes}}: We consider the following two incidents: (i) the active BS fails at time slot $n=120$ and (ii) the network suddenly experiences a connectivity problem and splits into two components (one component contains BS $1$ and BS $2$ and the other component contains BS $3$, BS $4$ and BS $5$) at time $n=120$. Because of the proposed architecture in Section~\ref{sec:recoverfromfaiulre}, RR and HEF are robust to the aforementioned incidents, and FIXED is not. We record the ``robustness'' of all these three schemes in Table~\ref{tab:comparision_different_scheduling_algorithms}.}
If we run the {\changed RR} algorithm, the remaining BSs will have the same active time, which is not necessarily optimal. In Figure~\ref{fig:adaptability_annotated}, we show the ratios of active time for all BSs in both considered scenarios. We see that the performance of HEF is always close to that of OPT before and after the network changes. {\changed Consequenlty, this shows that HEF reacts rapidly to network changes.}

{\bf{Communication overhead}}: Figure \ref{fig:communication_overhead} shows the overall number of packets transmitted per hour by using short-range communication when using different algorithms. FIXED only transmits data packets and does not need to exchange any other control messages. It serves as a baseline in the comparisons. HEF has additional packet exchanges of \texttt{BS\_ADVERT}, \texttt{BS\_UP} and \texttt{BS\_UP\_ACK} messages. Because these messages are sent at low rates, \textit{e.g.}, $1$ packet per $5$ minutes for \texttt{BS\_ADVERT} and $1$ packet every $2$ hours for \texttt{BS\_UP} and \texttt{BS\_UP\_ACK}, the communication overhead of HEF is almost negligible. The communication overhead of {\changed RR} is the same as HEF because they have the same amount of control messages. 
{\changed We summarize the result in the fourth column of Table~\ref{tab:comparision_different_scheduling_algorithms}.
}
\begin{table}[t]
\renewcommand{\arraystretch}{1.3}
\begin{center}
    \caption{Comparisons of Different Algorithms}
  \begin{tabular}{|c|c|c|c|} 
     \hline
     Algorithms  & Energy efficiency & Robustness& Overhead \\
    \hline
    FIXED & low &  {\changed no} & none   \\
    \hline
    {\changed RR} & medium &  {\changed yes} & low   \\
    \hline
     HEF & high & {\changed yes}  & low   \\
     \hline
     OPT & high & - & -\\
    \hline
  \end{tabular}
      \label{tab:comparision_different_scheduling_algorithms}
    \end{center}
\end{table}

\subsection{Validations of optimality conditions}
\label{subsec:validation_opt_condition}
\begin{figure}[t]
    \centering  
        \includegraphics[width=0.98\linewidth]{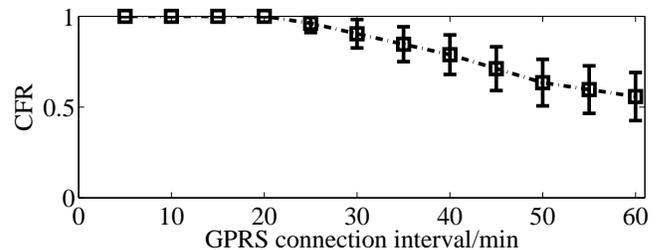}
    \caption{\changed  The condition fulfilled ratio (CFR) versus the GPRS connecting interval. Confidence intervals of $95\%$ are used. We see that condition D$4$ is always satisfied with a GPRS interval less than $20\,\rm{min}$.}
    \label{fig:gprssolarsieze}
\end{figure}

{
\changed 
In Theorem~\ref{thm:hef_opt}, we need conditions D$3$ and D$4$ to ensure the asymptotic optimality of HEF. In the following, we test the validity of these conditions.

Condition D$3$ requires that for any passive BS, the expected energy gain from the solar panel is larger than the energy consumption regardless of the selection of the active BS. It equals that the sizes of solar panels are large enough to support the operations for any  passive BSs. To validate condition D$3$, we generate $50$ sensor networks with the sensor nodes distributed in the sensing field uniformly at random. In Figure~\ref{fig:sensingsolarsize}, we show the average of the required sizes of solar panels in all these generated random networks under different data generating rates. Confidence intervals of $95\%$ are used. We see that condition D$3$ is easily satisfied: equipping all BSs with a $50\,\rm{cm}^2$ solar panel is enough when the data generating rates are less than $60\,\rm{packets}/\rm{min}$.

\begin{figure}[h]
    \centering  
        \includegraphics[width=0.98\linewidth]{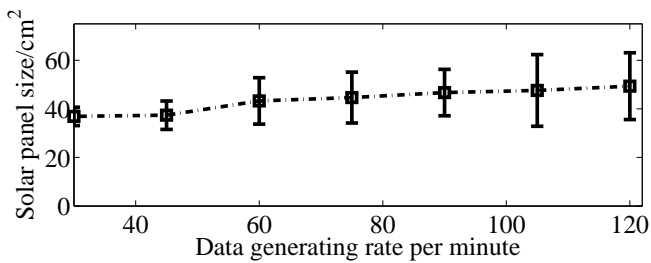}
    \caption{\changed The minimum size of solar panels required by condition D$3$ in Theorem~\ref{thm:hef_opt} under different data generating rates. Confidence intervals of $95\%$ are used. We see that the required size of solar panels slightly increases with the data generating rate. Equipping all BSs with a $50\,\rm{cm}^2$ solar panel is sufficient to satisfy condition D$3$ with a data generating rate at $60\,\rm{packets}/\rm{min}$. }
    \label{fig:sensingsolarsize}
\end{figure}

Condition D$4$ requires that the energy consumption rates of active BSs are much larger than the differences of energy consumption rates among all passive BSs. The energy consumption rates of active BSs mainly depend on the time interval between every two GPRS connections. The larger the GPRS connection interval, the smaller the energy consumption rates of active BSs. In Figure~\ref{fig:gprssolarsieze}, we randomly generate $50$ sensor networks and test the validity of condition D$4$ under different GPRS connection intervals. We define the \emph{condition fulfilled ratio} (CFR) as the fractions of instances that the generated sensor network fulfils condition D$4$. We see that condition D$4$ is always satisfied with a GPRS connection interval less than $20\,\rm{min}$.
}

\begin{figure}[h]
    \centering
    \includegraphics[width=\linewidth]{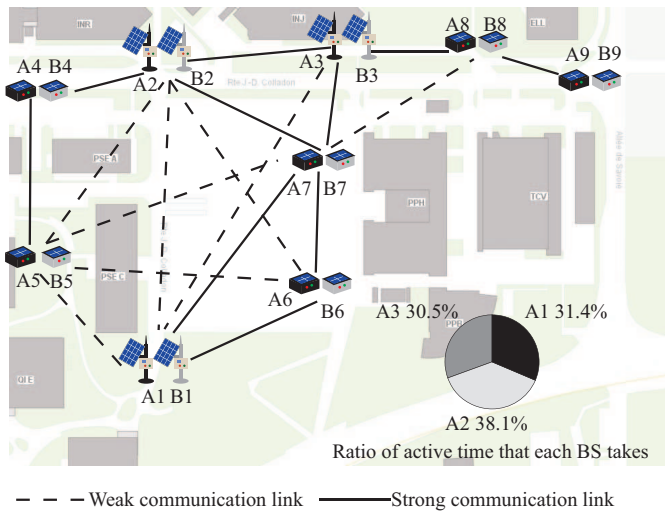}
    \caption{Experiment testbed on our campus. Two groups of $9$ sensor nodes are installed at the same locations. The two groups use different communication radio frequencies and thus form two separate networks. In the network with the black nodes, we use HEF to coordinate $3$ BSs, $A1$, $A2$ and $A3$. They are active for $31.4\%$, $38.1\%$ and $30.5\%$ of the total time respectively. The network with the white nodes has a fixed active BS $B2$. The solid lines represent communication links  of sustained good quality. The dotted lines represent temporary communication links.}
    \label{fig:experiement_setup}
\end{figure}

\begin{figure*}[t]
    \centering
   \includegraphics[width=.98\linewidth]{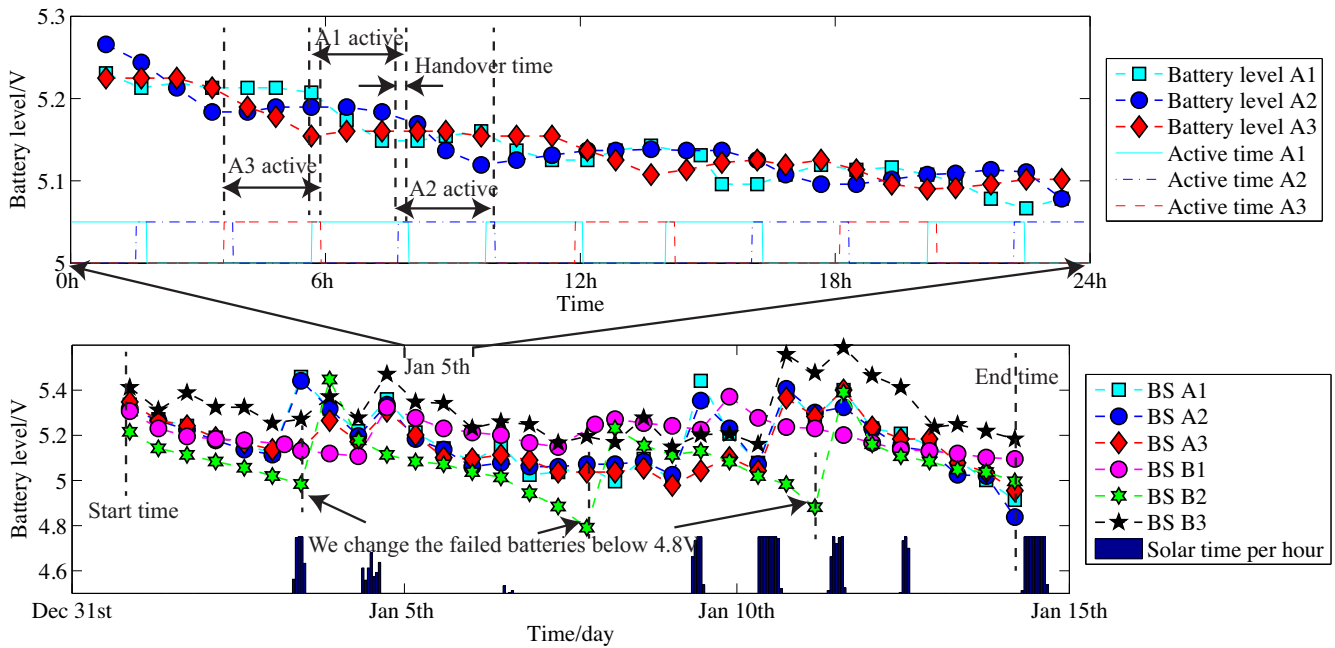}
    \caption{Battery levels of the six BSs in the real experiment versus time. BSs with ID $A_1$,$A_2$ and $A_3$ share the burden of being active and run the HEF algorithm. As a comparison, $B_2$ is a always-active BS while BSs $B_1$ and $B_3$ are always passive. We observe two facts as follows. First, the amounts of available energy of the BSs $A_1$, $A_2$ and $A_3$ are almost all the same during this $15$ days. To clarify this point, we especially investigate the data on Jan $5$th. We see that the active BS consumes energy quickly in each time interval of two hours. However, BSs running HEF take turns to share this high cost and averages out the temporal and spatial variations of the energy captured from solar panels. Second, by running the proposed scheme, the lifetime of the WSN is prolonged. We have to change $3$ times the batteries of BS $B_2$ on Jan $3$th, Jan $7$th and Jan $11$th.  Meanwhile, we do not need to change the batteries for the network running HEF during the whole $15$ days.}
    \label{fig:Jan_10th}
\end{figure*} 
 
\section{Real Experiments}
\label{sec:real_experiment}
We run a $15$-day experiment on an outdoor testbed on our campus based on the project Sensorscope~\cite{2010-TOSN-INGELREST}. As shown in Figure~\ref{fig:experiement_setup}, we deploy $2$ different networks at the same $9$ locations, resulting in a total number of $18$ sensor nodes. These two networks use separately $868\, \rm{MHz}$ and  $870\, \rm{MHz}$ frequency bands and thus do not interfere with each other. The general architecture of these two networks are the same as discussed in Section~\ref{sec:architecture}. The first network $N_1$ is composed of $3$ BSs ($A_1$, $A_2$ and $A_3$) and $6$ regular sensor nodes ($A_4$, $A_5$, $A_6$, $A_7$, $A_8$ and $A_9$). This network runs HEF to adaptively choose one active BS. The second network $N_2$ is also composed of $3$ BSs ($B_1$, $B_2$ and $B_3$) and $6$ regular sensor nodes ($B_4$, $B_5$, $B_6$, $B_7$, $B_8$ and $B_9$). It runs FIXED, which keeps BS $B_2$ active and BSs $B_1$, $B_3$ passive. 
The data packet is generated as follows. Each sensor node generates a $2$-byte counter every $30\, \rm{sec}$. The value of the counter changes according to a triangular waveform. Then, each counter is attached with a $4$-byte timestamp and a $2$-byte indicator for indicating message types. We duplicate them into four copies and then encapsulate them into data packets. Each data packet has a $3$-byte header containing the node IDs and the hop distances to the active BS. The average data generating rate of each sensor node is $35\, \rm{byte}/ {30}\, \rm{sec}$. All these data packets are routed to the active BS who connects to the remote server by using GSM/GPRS every $5\, \rm{min}$. 
On average, the active BS transmits $9\times 35\,\rm{byte} \times 5\, \rm{min}/ 30\, \rm{sec}=3150 \, \rm{byte}$ data every $5\, \rm{min}$. 



In the experiment, we use the battery level as the indicator for the available energy. Every $5\, \rm{min}$, each BS sends its battery level to the active BS in a \texttt{BS\_ADVERT} message (Section~\ref{sec:gatherinfo}). The active BS then forwards this message to the remote server, hence we are able to observe the variations of the available energy in the WSN. Notice that this message is transmitted with a low rate and it will not add much communication burden to the network. 

BSs and regular sensor nodes are equipped with solar panels with areas of $100\, \rm{cm}^2$  and $50\, \rm{cm}^2$, respectively. 
They are all equipped with $4$ AA NiMH rechargeable batteries (each battery has a capacity of $800\, \rm{mAh}$). In Figure~\ref{fig:Jan_10th}, we show the battery levels of the six BSs. We see that in network $N_1$,  the $3$ BSs with ID $A_1$, $A_2$ and $A_3$ almost always keep the same battery levels, although their solar panels harvest different amounts of energy. 
During this period of $15$ days, their batteries do not deplete. Meanwhile, in network $N_2$, the passive BSs $B_1$ and $B_3$ always have high battery levels because of their low energy consumption rates. The always-active BS  $B_2$ consumes its battery quickly and on the $4$th, $8$th and $12$th days, the batteries of $B_2$ drain out and we have to change them. From the experiment, we conclude that by deploying multiple BSs and adaptively choosing the active BS, the harvested energy is fully used and the network lifetime is prolonged.

\section{Conclusion}
\label{sec:conclusion}
In this paper, we have presented and evaluated a novel scheme for organizing WSNs, in which multiple BSs are deployed but only one BS is adaptively selected to be active. By using the proposed scheme, we efficiently utilize the temporally and spatially varying energy resources available to all BSs. Therefore, the large batteries and energy harvesting devices of individual BSs can be substantially reduced.

To adaptively choose the active BS, we have proposed a simple yet powerful algorithm HEF. We have proved its asymptotic optimality under mild conditions. 

Through simulations on the simulator Omnet++/Castalia and real experiments on an outdoor testbed, we have shown that the proposed scheme is energy-efficient, is adaptable to network changes and is low in communication overhead.

{
\changed 
In future work, we intend to investigate the scheme where multiple BSs are allowed to be simultaneously active in a very large WSN. 
In this scheme, we have to design new algorithms to adaptively select a group of active BSs, and jointly optimize the energy efficiency of both BSs and regular sensor nodes. 
Many new implementation issues need to be tackled, for example: (i) bootstrapping the network into a steady state with multiple active BSs and (ii) handover of the roles of active BSs from a group of BSs to another group of BSs.
}


\section*{Acknowledgements}
The work presented in this paper was supported (in part) by the Swiss National Science Foundation under grant number 200021-146423. {\changed We would like to thank Dr. Olivier Leveque for useful discussions and insightful suggestions on the proofs of Theorems~\ref{lm:relation_between_opts} and~\ref{thm:hef_opt}.}


\ifCLASSOPTIONcaptionsoff
  \newpage
\fi

{\footnotesize
\bibliographystyle{abbrv}
\bibliography{sensor-networks}
}

%
%

%
%

\clearpage
\clearpage

\appendices
\section{Azuma-Hoeffding inequality}
In both the proofs of Theorem~\ref{lm:relation_between_opts} and Theorem~\ref{thm:hef_opt}, we will use the Azuma-Hoeffding inequality~\cite[p. 476]{Geoffrey01} for martingales with bounded differences. We repeat it here for convenience.
\begin{lemma}
Suppose $\{H^{(n)}\}_{n \in \mathbb{N}}$ is a martingale and
\(|H^{(n)} - H^{(n-1)}| < c_n\)
almost surely, where $c_n$ is positive real for any $n \in \mathbb{N}$. Then for any positive integer $N$ and any positive $\gamma$, 
\[
P(H^{(N)} - H^{(0)} \geq \gamma) \leq \exp\left ({-\gamma^2 \over 2 \sum_{n=1}^N c_n^2} \right),
\]
and conversely
\[
P(H^{(N)} - H^{(0)} \leq -\gamma) \leq \exp\left ({-\gamma^2 \over 2 \sum_{n=1}^N c_n^2} \right).
\]
\label{lm:azuma_hoeffding}
\end{lemma}

In our sensor network scenario, for any BS $1\leq m \leq M$, we construct a martingale $\{{h}_m^{(n)}\}_{n \in \mathbb{N}}$ with
\begin{equation}
h_m^{(n)}= e_m^{(0)}+ \tau \sum_{t=1}^n \left({s}_m^{(t)}- \bar{s}\right),
\label{eq:def_hm}
\end{equation}
where
\[
\mathbb{E}_{n-1} {h}_m^{(n)}= {h}_m^{(n-1)}
\]
because of (\ref{eq:recharge_rate_constant_expectation}).
Using (\ref{eq:sum_iterative_function}) and (\ref{eq:defR}), we have
\[
e_m^{(n)}= e_m^{(0)}+ \tau \sum_{t=1}^n \left({s}_m^{(t)}- \bar{s}\right) -\tau \sum_{t=1}^{n} (\bm{R}\bm{v}^{(t)})_m.
\]
Therefore, (\ref{eq:def_hm}) is equivalent to 
\begin{equation}
h_m^{(n)}= e_m^{(n)}+ \tau \sum_{t=1}^{n} (\bm{R}\bm{v}^{(t)})_m.
\label{eq:relationhmem}
\end{equation}
Using Lemma~\ref{lm:azuma_hoeffding}, we have the following result.
\begin{corollary}
When the energy recharge rates $\{\bm{s}^{(n)}\}_{n \in \mathbb{N}}$ satisfy conditions (\ref{eq:recharge_rate_constant_expectation}) and (\ref{eq:recharge_rate_boundedness}), for any $n_1<n_2\in \mathbb{N}$ and for any $\Delta_1,\Delta_2>0$,   
\begin{align}
\nonumber & \mathbb{P}\left(h_m^{(n_2)}- h_m^{(n_1)}\leq - \Delta_1 -(n_2-n_1)\Delta_2 \right) \\
\nonumber &  \qquad \leq  \exp \left(\frac{-( \Delta_1 +(n_2-n_1)\Delta_2)^2}{2(n_2-n_1)\tau^2 S^2 }\right)\\
\nonumber & \qquad \leq  \exp \left(\frac{-(n_2-n_1) \Delta_2^2}{2\tau^2S^2} \right) \cdot \exp \left(-\frac{\Delta_1\Delta_2}{\tau^2 S^2} \right),
\end{align}
because of (\ref{eq:def_hm}) and because (\ref{eq:recharge_rate_boundedness}) yields that $s_m^{(n)} \leq S$.
Similarly
\begin{align}
\nonumber & \mathbb{P}\left(h_m^{(n_2)}- h_m^{(n_1)}\geq  \Delta_1 +(n_2-n_1)\Delta_2 \right) \\
\nonumber & \qquad \leq  \exp \left(\frac{-(n_2-n_1) \Delta_2^2}{2\tau^2S^2} \right) \cdot \exp \left(-\frac{\Delta_1\Delta_2}{\tau^2 S^2} \right).
\end{align}
\label{coro:wsnazu}
\end{corollary}

\section{Proof of Theorem \ref{lm:relation_between_opts}}
\label{ap:proofoflemmarelation}
(i) We first show that (\ref{eq:thm1r2})  holds when $f^*<0$. We consider a scheduling algorithm that lets the cumulative active time for BS $m$ at time slot $n$ be
\[
\sum_{t=1}^n {v}_m^{(t)}=
\left\{
\begin{aligned}
&\lfloor n\bar{v}_m^* \rfloor, &&\  1\leq m \leq M-1,\\
& n- \sum_{k=1}^{M-1}\lfloor n\bar{v}_k^* \rfloor, &&\  m = M,
\end{aligned}
\right.
\]
where $\bar{v}_m^*$  is the $m$-th element of the optimal solution $\bm{\bar{v}}^*$ of problem (\ref{eq:max_dropping_rate}). We see that: for any $1\leq m \leq M-1$, $n {\bar{v}}_m^* -1\leq  \sum_{t=1}^n {v}_m^{(t)} \leq  n {\bar{v}}_m^*$; for $m=M$,  $n {\bar{v}}_m^* \leq  \sum_{t=1}^n {v}_m^{(t)} \leq  n {\bar{v}}_m^*+M-1$ due to $\sum_{m=1}^M n {\bar{v}}_m^*=n$ and $n {\bar{v}}_k^*-1 \leq \lfloor n\bar{v}_k^* \rfloor \leq n {\bar{v}}_k^*$ for $1\leq k\leq M-1$. Consequently, 
\begin{equation}
\left\| n \bm{\bar{v}}^*-\sum_{t=1}^n \bm{v}^{(t)} \right\|_{\infty }< M.
\label{decionsumvets}
\end{equation}
We define the lifetime of each BS $m$ ($1\leq m\leq M$) as
\begin{align*}
N_{m}= \inf \{\{\infty\}\cup \{n | e_m^{(n+1)}<0 ,  n \in \mathbb{N}\}\}.
\end{align*}
The lifetime of the whole network using the aforementioned scheduling algorithm is 
\begin{equation*}
N^*= \min_{1\leq m\leq M}N_{m}.
\end{equation*}
Because $N_{\rm{opt}}$ is the optimal lifetime, we have $N^* \leq N_{\rm{opt}}$. To show (\ref{eq:thm1r2}), it suffices to show that
\begin{equation}
\lim_{e_0 \rightarrow \infty} \mathbb{P}  (N^*=\infty)= 1.
\label{eq:eocongergsmatbot}
\end{equation}
Because of (\ref{eq:relationhmem}), $e_m^{(n)} <0$ implies that
\begin{align}
\nonumber h_m^{(n)} &< \tau \sum_{t=1}^n  (\bm{R}\bm{v}^{(t)})_m\\
\nonumber & = n \tau(\bm{R} \bm{\bar{v}}^*)_m +   \tau\bm{R}  (\sum_{t=1}^n \bm{v}^{(t)}- n\bm{\bar{v}}^*)_m \\
\nonumber &<n \tau(\bm{R} \bm{\bar{v}}^*)_m+M\tau  \left\| \bm{R} \right\|_{\infty} \cdot \left\|\sum_{t=1}^n \bm{v}^{(t)}- n\bm{\bar{v}}^* \right\|_{\infty}\\
&<n \tau a_m+ M^2\tau \|\bm{R}\|_{\infty}, 
\label{eq:hmdeieifjijfsem}
\end{align}
where in the second inequality we use  (\ref{decionsumvets}) and in the third inequality we define $a_m= (\bm{R} \bar{\bm{v}}^*)_m$. 
Due to $\bm{R} \bm{\bar{v}}^* \leq f^* \bm{u}_M$ in problem (\ref{eq:max_dropping_rate}), we have $a_m\leq f^*<0$.

Using the union bound, we have 
\begin{align}
\nonumber & \mathbb{P}  \left(  N_{m} < \infty\right) \\
\nonumber & \qquad = \mathbb{P}(\exists n \geq 1: e_m^{(n+1)} <0)\\
\nonumber & \qquad = \mathbb{P}(\exists n \geq 2: e_m^{(n)} <0)\\
\nonumber &\qquad \leq \mathbb{P}(\exists n \geq 2: h_m^{(n)} <n\tau a_m+M^2\tau \|\bm{R}\|_{\infty})\\
&\qquad  \leq \sum_{n=2}^{\infty} \mathbb{P}(h_m^{(n)}< n\tau a_m+M^2\tau \|\bm{R}\|_{\infty})
\label{eq:tmseparate}
\end{align}
Using Corollary~\ref{coro:wsnazu}, we have 
\begin{align}
\nonumber &\mathbb{P}(h_m^{(n)}<n \tau a_m+M^2\tau \|\bm{R}\|_{\infty}) \\
\nonumber & \qquad= \mathbb{P}\left(h_m^{(n)}- h_m^{(0)}< -e_0+ n\tau a_m+ M^2\tau \|\bm{R}\|_{\infty}\right)\\
& \qquad \leq \exp \left(\frac{-n a_m^2}{2S^2} \right) \cdot \exp \left(\frac{(e_0-M^2\tau \|\bm{R}\|_{\infty})  a_m}{\tau S^2} \right)
\label{eq:hmexpam}
\end{align}
By using (\ref{eq:tmseparate}) and (\ref{eq:hmexpam}) together and by noticing that $\sum_{n=2}^\infty \exp \left({-na_m^2}/{2S^2} \right)=\beta<\infty$,
we have that
\[
\mathbb{P}  \left(  N_{m}<\infty\right)<\beta \exp \left(\frac{(e_0-M^2\tau \|\bm{R}\|_{\infty}) a_m}{\tau S^2}\right).
\]
Using the union bound and taking the limit $e_0 \rightarrow \infty$, we have
\[
\lim_{e_0 \rightarrow \infty} \mathbb{P}  (N^*<\infty)\leq \lim_{e_0 \rightarrow \infty}  \sum_{m=1}^M  \mathbb{P} (N_m < \infty)=0,
\]
which implies (\ref{eq:eocongergsmatbot}). Therefore, we have (\ref{eq:thm1r2}).

(ii) Then, we will show that (\ref{eq:thm1r1}) holds when $f^*>0$ in the following two separate parts:
\begin{empheq}[left=\empheqlbrace]{align}
\label{eq:noptsmale}
\lim_{e_0 \rightarrow \infty} \mathbb{P} \left({N_{\rm{opt}}}\leq (1-\delta){e_0/(\tau f^*)} \right)=0, \forall \delta>0,\\
\label{eq:noptlarge} \lim_{e_0 \rightarrow \infty} \mathbb{P} \left( {N_{\rm{opt}}}\geq (1+\delta){e_0/(\tau f^*)} \right)=0, \forall \delta>0.
\end{empheq}

1) In the first part, to show (\ref{eq:noptsmale}), it suffices to show that
\begin{equation}
\lim_{e_0 \rightarrow \infty} \mathbb{P} (N^*\leq (1-\delta){e_0/(\tau f^*)}) =0, \forall \delta>0,
\label{eq:ptstatsoee}
\end{equation}
because $N^*\leq N_{\rm{opt}}$.


Because of (\ref{eq:hmdeieifjijfsem}), $e_m^{(n+1)} <0$ implies that $h_m^{(n+1)} < (n+1)\tau a_m+ M^2\tau \|\bm{R}\|_{\infty}$.
Using the union bound,  
\begin{align}
\nonumber & \mathbb{P}  \left(  N_{m} \leq  \frac{(1-\delta)e_0}{\tau f^*} \right) \\
\nonumber & \qquad = \mathbb{P}\left( \exists n \leq \frac{(1-\delta)e_0}{\tau f^*}: e_m^{(n+1)} <0\right)\\
\nonumber & \qquad = \mathbb{P}\left( \exists n \leq \frac{(1-\delta)e_0}{\tau f^*}+1: h_m^{(n)}< n\tau a_m+M^2\tau \|\bm{R}\|_{\infty} \right)\\
 &\qquad \leq \sum_{n=1}^{{(1-\delta)e_0}/{\tau f^*}+1} \mathbb{P}\left(  h_m^{(n)}< n\tau a_m+M^2\tau \|\bm{R}\|_{\infty}\right).
\label{eq:tmseparate2}
\end{align}
Now, using Lemma~\ref{lm:azuma_hoeffding}, we have 
\begin{align}
\nonumber &\mathbb{P}(h_m^{(n)}<n \tau a_m+M^2\tau \|\bm{R}\|_{\infty}) \\
\nonumber & \qquad =\mathbb{P}\left(h_m^{(n)}- h_m^{(0)}< -e_0+M^2\tau \|\bm{R}\|_{\infty}+ n \tau a_m \right)\\
\nonumber & \qquad \leq  \exp \left(\frac{-(-e_0+ M^2\tau \|\bm{R}\|_{\infty}+n\tau a_m)^2}{2n\tau^2S^2 }\right)\\
 & \qquad \leq \exp \left(\frac{-(\delta e_0-M^2\tau \|\bm{R}\|_{\infty})^2}{2(1-\delta)e_0\tau S^2/ f^*} \right),
\label{eq:hmnsdatam}
\end{align}
where the inequality on the fourth line is due to $n\leq (1-\delta) e_0 /(\tau f^*)$, and $a_m \leq f^*$ which follows from $a_m= (\bm{R} \bar{\bm{v}}^*)_m$ and $\bm{R} \bm{\bar{v}}^* \leq f^* \bm{u}_M$.

Combining (\ref{eq:tmseparate2}) and (\ref{eq:hmnsdatam}) yields that 
\begin{align}
\nonumber & \mathbb{P}  \left(  N_{m} \leq \frac{(1-\delta)e_0}{\tau f^*} \right)  \\
& \qquad \leq \left(\frac{(1-\delta)e_0}{\tau f^*}+1\right) \exp \left(\frac{-(\delta e_0-M^2\tau \|\bm{R}\|_{\infty})^2  f^*}{2(1-\delta)e_0\tau S^2} \right).
\label{eq:tdeldfstafsf}
\end{align}
Using the union bound, we see that
\[
\mathbb{P}  \left(  N^*\leq  \frac{(1-\delta)e_0}{\tau f^*} \right) \leq \sum_{m=1}^M \mathbb{P}  \left(  N_{m}\leq \frac{(1-\delta)e_0}{\tau f^*} \right).
\]
Plugging in (\ref{eq:tdeldfstafsf}) and taking $e_0\rightarrow \infty$, we have (\ref{eq:ptstatsoee}).

2) In the second part, we will show (\ref{eq:noptlarge}). Consider the event that $N_{\rm{opt}} \geq (1+\delta){e_0/(\tau f^*)}$. It implies that there exists a sequence of decision vectors $\{\bm{v}^{(t)}\}_{t\in \mathbb{N}}$  such that the available energy at time $(1+\delta){e_0/(\tau f^*)}$ is non-negative, \textit{i.e.}, 
\begin{equation}
\bm{e}^{( \frac {(1+\delta)e_0}{\tau f^*})}\geq \bm{0}.
\label{eq:awerne1pedel}
\end{equation}
Let $\bar{\bm{v}}_{\rm{opt}}$ be the average decision vector from time $1$ to $(1+\delta){e_0/(\tau f^*)}$, that is,
\begin{equation}
\bar{\bm{v}}_{\rm{opt}}=\frac{1}{(1+\delta){e_0/(\tau f^*)}} \sum_{n=1}^{(1+\delta){e_0/(\tau f^*)}} \bm{v}^{(n)}.
\label{eq:defvoptsdfs}
\end{equation}
Using (\ref{eq:sum_iterative_function}), (\ref{eq:awerne1pedel}) and that $\bm{e}^{(0)}= e_0 \bm{u}_M$, we have that
\begin{align}
\nonumber e_0 \bm{u}_M+ \sum_{n=1}^{ (1+\delta){e_0/(\tau f^*)}} \tau \bm{s}^{(n)} \geq \tau \bm{C}\sum_{n=1}^{(1+\delta){e_0/(\tau f^*)}} \bm{v}^{(n)}.
 \label{eq:avaioneplusdelta}
\end{align}
Substracting $(1+\delta){e_0/(\tau f^*)} \bm{\bar{s}}$ on both sides of this inequality and using (\ref{eq:defR}) and (\ref{eq:defvoptsdfs}), we have that 
\begin{align}
e_0 \bm{u}_M+\tau \sum_{n=1}^{ (1+\delta){e_0/(\tau f^*)}} (\bm{s}^{(n)}- \bm{\bar{s}}) \geq  \frac{(1+\delta)e_0}{f^*} \bm{R}  \bar{\bm{v}}_{\rm{opt}}.
\end{align}
Because $f^*$ is the optimal objective value of (\ref{eq:max_dropping_rate}), there exists $1\leq m^* \leq M$, such that  $(\bm{R} \bm{\bar{v}}_{\rm{opt}} )_{m^*} \geq f^*$. Plugging it into (\ref{eq:avaioneplusdelta}) and dividing both sides of the equation by $(1+\delta)e_0/ f^*$, we have
\begin{equation}
\frac{1}{(1+\delta){e_0/(\tau f^*)}} \sum_{t=1}^{ (1+\delta){e_0/(\tau f^*)}} ({s}_{m^*}^{(t)}- \bar{s}_{m^*})  \geq \frac{f^*\delta}{(1+\delta)}.
\label{eq:eventlargerpsdof}
\end{equation} 
Because ${s}_{m^*}^{(t)}- \bar{s}_{m^*}$ is a martingale difference term, using the weak law of large numbers, $\sum_{n=1}^{n_0}  ({s}_{m^*}^{(n)}- \bar{s}_{m^*}) /n_0$ weakly converges to $0$ when $n_0\rightarrow \infty$. Therefore, the probability of (\ref{eq:eventlargerpsdof}) converges to zero when $e_0 \rightarrow \infty$. Remember that event (\ref{eq:eventlargerpsdof}) is implied by $N_{\rm{opt}} \geq (1+\delta){e_0/(\tau f^*)}$,  which concludes the proof.  

\section{Proof of Theorem \ref{thm:hef_opt}}
\label{ap:proofoflemmarelation}
\subsection{Notations and a short summary}
For the ease of discussion, we define four events:
\begin{align}
\nonumber &A_1= \{\exists N_{\rm{hef}}\leq \frac{(1-\delta)e_0}{\tau f^*}, l^* \in [1, M], \textit{s.t.}, e_{l^*}^{(N_{\rm{hef}}+1)}<0\},\\
\nonumber&A_1'=\{\exists N_{\rm{hef}}\leq Ke_0,  l^* \in [1, M], \textit{s.t.}, e_{l^*}^{(N_{\rm{hef}}+1)}<0\},\\
\nonumber &A_2=\{e_{m}^{(N_{\rm{hef}}+1)}<\epsilon_1 e_0, \forall 1\leq m \leq M\}, \\
\label{eq:def_eventa3}  &A_3= \left\{\left\| \frac{1}{N_{\rm{hef}}+1}\sum_{n=1}^{N_{\rm{hef}}+1} \bm{v}^{(n)} - \frac{\bm{R}^{-1} \bm{u}_M  }{\bm{u}_M^{\top}  \bm{R}^{-1} \bm{u}_M } \right\|_{\infty}<\epsilon_2 \right\},
\end{align}
where $K$ is the constant defined in (\ref{eq:thmresult2}) and $\epsilon_1,\epsilon_2$ are two positive numbers which can be set arbitrarily small. 

Events $A_1$ and $A_1'$  mean that the lifetime incurred by HEF $N_{\rm{hef}}$ is not larger than ${(1-\delta)e_0}/{\tau f^*}$ and $Ke_0$, respectively. 
Event $A_2$ occurs when the available energy of any BS is smaller than $\epsilon_1 e_0$ at time $N_{\rm{hef}}+1$. 
Event $A_3$ means that the average of decision vectors up to time $N_{\rm{hef}}+1$ is arbitrarily close to ${\bm{R}^{-1} \bm{u}_M  }/{\bm{u}_M^{\top}  \bm{R}^{-1} \bm{u}_M }$.

Because $N_{\rm{hef}} \leq N_{\rm{opt}}$ and because of (\ref{eq:thm1r1}), when $f^*>0$,
\begin{equation}
\lim_{e_0 \rightarrow \infty} \mathbb{P} \left(   \frac{N_{\rm{hef}}}{e_0/(\tau f^*)}-1<\delta \right)=1, \forall \delta>0.
\end{equation}
Therefore, Theorem~\ref{thm:hef_opt} can be recast as: 
\begin{align}
\label{eq:resulta1psdfier}
&\lim_{e_0 \rightarrow \infty} \mathbb{P} (A_1)=0, \mbox{ if } f^*>0,\\
& \lim_{e_0 \rightarrow \infty} \mathbb{P} (A_1')=0, \mbox{ if } f^*<0.
\label{eq:resulta2psdfier}
\end{align} 

First,  using Azuma-Hoeffding inequality~\cite[p. 476]{Geoffrey01} and condition D$3$, we will show that the probability that HEF uses up all the available energy of all BSs converges to $1$ when $e_0\rightarrow \infty$. More precisely, we will show that 
$\lim_{e_0 \rightarrow \infty} \mathbb{P} (A_1' \cap \bar{A}_2)=0$ and $\lim_{e_0 \rightarrow \infty} \mathbb{P} (A_1 \cap \bar{A}_2)=0$.
 
Secondly, using the duality of linear programming and condition D$4$, we will show that ${\bm{R}^{-1} \bm{u}_M  }/{\bm{u}_M^{\top}  \bm{R}^{-1} \bm{u}_M }$ is the optimal fractions of active time for all BSs. Then, we will show that if event $A_3$ is true, HEF performs optimally. We will deduce that: if $f^*<0$, $\lim_{e_0 \rightarrow \infty} \mathbb{P} (A_1' \cap A_3)=0$; and if $f^*>0$, $\lim_{e_0 \rightarrow \infty} \mathbb{P} (A_1 \cap A_3)=0$.

Thirdly, we will show that if HEF uses up the available energy of all BSs (event $A_2$ is true), the average of decision vector $\sum_{n=1}^{N_{\rm{hef}}+1} \bm{v}^{(n)}/({N_{\rm{hef}}}+1)$ is close to ${\bm{R}^{-1} \bm{u}_M  }/{\bm{u}_M^{\top}  \bm{R}^{-1} \bm{u}_M }$ (event $A_3$ is true). Then, we will deduce that: if $\lim_{e_0 \rightarrow \infty} \mathbb{P} (A_1' \cap A_3)=0$, we have $\lim_{e_0 \rightarrow \infty} \mathbb{P} (A_1' \cap A_2)=0$; and if $\lim_{e_0 \rightarrow \infty} \mathbb{P} (A_1 \cap A_3)=0$, we have $\lim_{e_0 \rightarrow \infty} \mathbb{P} (A_1 \cap A_2)=0$.

To sum things up, to prove (\ref{eq:resulta1psdfier}) and (\ref{eq:resulta2psdfier}), we will show the following six separate points:
\begin{itemize}
\item{Point $1A$:} $\lim_{e_0 \rightarrow \infty} \mathbb{P} (A_1' \cap \bar{A}_2)=0$.
\item{Point $1B$:} $\lim_{e_0 \rightarrow \infty} \mathbb{P} (A_1 \cap \bar{A}_2)=0$.
\item{Point $2A$:} If $f^*<0$, we have $\lim_{e_0 \rightarrow \infty} \mathbb{P} (A_1' \cap A_3)=0$.
\item{Point $2B$:} If $f^*>0$, we have $\lim_{e_0 \rightarrow \infty} \mathbb{P} (A_1 \cap A_3)=0$.
\item{Point $3A$:} Given that $\lim_{e_0 \rightarrow \infty} \mathbb{P} (A_1' \cap A_3)=0$, we have $\lim_{e_0 \rightarrow \infty} \mathbb{P} (A_1' \cap A_2)=0$.
\item{Point $3B$:} Given that $\lim_{e_0 \rightarrow \infty} \mathbb{P} (A_1 \cap A_3)=0$, we have $\lim_{e_0 \rightarrow \infty} \mathbb{P} (A_1 \cap A_2)=0$.
\end{itemize}
Then, combining Points $1A$, $2A$ and $3A$, we have 
\[
\lim_{e_0 \rightarrow \infty} \mathbb{P} (A_1')= \lim_{e_0 \rightarrow \infty} (\mathbb{P} (A_1' \cap A_2) + \mathbb{P} (A_1' \cap \bar{A}_2))=0,
\]
which proves (\ref{eq:resulta2psdfier}), and likewise, Points $1B$, $2B$ and $3B$ yield (\ref{eq:resulta1psdfier}).

For the ease of discussion, we define two constants 
\begin{empheq}[left=\empheqlbrace]{align}
\label{eq:def_d1} d_1= \max_{m \neq {j}} R_{mj}<0,\\
\label{eq:def_d2} d_2= \max_{m,j} C_{mj}> 0.
\end{empheq}

\subsection{Proof for point $1A$ and point $1B$}
We only show the proof for point $1A$ here. The proof for point $1B$ is identical if we replace $K$ with $(1-\delta) /(\tau f^*) $.
Let $l^*$ be a BS that drains out of energy at time $N_{\rm{hef}}+1$, \textit{i.e.}, $e_{l^*}^{(N_{\rm{hef}}+1)} < 0$.
Let $N'$ be the last time that BS $l^*$ is selected as the active BS (we set $N'=0$ if BS $l^*$ is never selected to be active), that is,
\[
N'=\sup \{\{0\}\cup \{n \mid \exists l^*, v_{l^*}^{(n)}=1, n\leq N_{\rm{hef}}\}\}.
\]
We define the event $A_4$ as 
\[
A_4= \{N_{\rm{hef}} -N'+1\geq  \epsilon_3 e_0\}.
\]
Using (\ref{eq:relationhmem}), we see that
\begin{align}
\nonumber h_{l^*}^{(N_{\rm{hef}}+1)}-h_{l^*}^{(N')}  & = e_{l^*}^{(N_{\rm{hef}}+1)}- e_{l^*}^{(N')}+ \tau \sum_{n=N'+1}^{N_{\rm{hef}}+1} (\bm{R} \bm{v}^{(n)})_{l^*} \\
 & < d_1 \tau (N_{\rm{hef}} -N'+1),
\label{eq:hlhefgelsmd}
\end{align}
where the inequality holds because $e_{l^*}^{(N_{\rm{hef}}+1)} < 0\leq e_{l^*}^{(N')}$, because BS $l^*$ is not selected as the active BS from time $N'+1$ to $N_{\rm{hef}}+1$ and because of (\ref{eq:def_d1}).

Using Corollary~\ref{coro:wsnazu}, we see that
\begin{align}
\nonumber & \mathbb{P}\left(h_{l^*}^{(N_{\rm{hef}}+1)}-h_{l^*}^{(N')} <  d_1 \tau (N_{\rm{hef}} -N'+1) \right)\\
 & \qquad \leq  \exp \left(\frac{-(N_{\rm{hef}}-N'+1) d_1^2}{2S^2} \right).
\label{eq:hmexpam2}
\end{align}
Using the union bound, we have
\begin{align}
\nonumber & \mathbb{P}(A_1' \cap A_4 ) \\
\nonumber & \quad \leq  \sum_{N'=1}^{Ke_0} \sum_{N_{\rm{hef}}=1}^{Ke_0} \mathbb{P}\left(h_{l^*}^{(N_{\rm{hef}}+1)}-h_{l^*}^{(N')} < d_1 \tau (N_{\rm{hef}} -N'+1)\right) \\
\nonumber & \quad \leq  (Ke_0)^2 \exp \left(-\frac{ (N_{\rm{hef}} -N'+1)   d_1^2}{2 S^2} \right) \\
& \quad \leq  (Ke_0)^2 \exp \left(-\frac{\epsilon_3 e_0  d_1^2}{2 S^2} \right),
\label{eq:aoneprimesppwewr}
\end{align}
where the second inequality follows from $N', N_{\rm{hef}} \leq Ke_0$ and (\ref{eq:hmexpam2}), and the third inequality follows from $N_{\rm{hef}} -N'+1\geq  \epsilon_3 e_0$ in the definition of $A_4$.

Taking $e_0 \rightarrow \infty$ on both sides of (\ref{eq:aoneprimesppwewr}), we see that $\lim_{e_0 \rightarrow \infty}\mathbb{P} (A_1' \cap A_4 ) =0$.

In the following, we will show that $\bar{A}_4  \subseteq A_2$. If $\bar{A}_4$ is true, $N_{\rm{hef}}  -N'+1<\epsilon_3 e_0$. Then, by summing up (\ref{eq:iterative_function}) from time $N'$ to $N_{\rm{hef}}+1$, we derive the following upper-bound for the available energy $e_{l^*}^{(N' )}$ 
\begin{align}
\nonumber e_{l^*}^{(N' )} &= e_{l^*}^{(N_{\rm{hef}}+1 )}- \sum_{n=N'+1}^{N_{\rm{hef}} +1} s_{l^*}^{(n)}+\sum_{n=N'+1}^{N_{\rm{hef}} +1} (\bm{C} \bm{v}^{(n)})_{l^*},\\
\nonumber &< \sum_{n=N'+1}^{N_{\rm{hef}} +1} (\bm{C} \bm{v}^{(n)})_{l^*}\\
& < d_2 \epsilon_3 e_0,
\end{align}
where the first inequality holds because $e_{l^*}^{(N_{\rm{hef}} +1)}<0$ and because $s_{l^*}^{(n)}>0$ for all $n$, and the second inequality holds because $N_{\rm{hef}}  -N'+1<\epsilon_3 e_0$, because $l^*$ is not selected as the active BS between time $N'+1$ to time $N_{\rm{hef}}+1$ and  because of (\ref{eq:def_d2}).

Because BS $l^*$ is selected by HEF at time $N'$, it has the highest available energy among all BSs. Therefore, we have $e_{m}^{(N')}<d_2 \epsilon_3 e_0$ for all $1\leq m \leq M$. This leads to the following upper-bound for the available energy of all BSs at time $N_{\rm{hef}}+1$,
\begin{align}
\nonumber e_{m}^{(N_{\rm{hef}}+1)} & = e_{m}^{(N')} + \sum_{n=N'+1}^{N_{\rm{hef}}+1} s_{m}^{(n)} - \sum_{n=N'+1}^{N_{\rm{hef}}+1} (\bm{C} \bm{v}^{(n)})_m \\
\nonumber & <  d_2 \epsilon_3 e_0+ (N_{\rm{hef}}-N'+1)S \\ 
& < d_2 \epsilon_3 e_0+ \epsilon_3 e_0 S = \epsilon_1 e_0.
\end{align}
where in the first inequality we use  $0\leq s_{m}^{(n)} \leq S$ and in the second inequality we use  $N_{\rm{hef}} -N'+1<  \epsilon_3 e_0$ and set $\epsilon_1= (d_2+S) \epsilon_3$.

Therefore, if $\bar{A}_4$ holds,  so does $A_2$. Hence, $\bar{A}_4  \subseteq A_2$ which follows that $\bar{A}_2 \subseteq A_4$. Consequently, $A_1' \cap \bar{A}_2 \subseteq A_1' \cap {A}_4$. It follows that
$\lim_{e_0 \rightarrow \infty} \mathbb{P} (A_1' \cap \bar{A}_2 ) \leq \lim_{e_0 \rightarrow \infty} \mathbb{P} (A_1' \cap A_4 ) =0$.

\subsection{Proof for point $2A$ and point $2B$}
We consider the following two scenarios when showing both points $2A$ and $2B$: (i) neither  $\bm{R}^{-1}\bm{u}_M \geq \bm{0}$ nor $\bm{R}^{-1}\bm{u}_M \leq \bm{0}$ is satisfied; (ii) either $\bm{R}^{-1}\bm{u}_M \geq \bm{0}$ or $\bm{R}^{-1}\bm{u}_M \leq \bm{0}$ is satisfied.

(i) In the first scenario, the vector ${\bm{R}^{-1} \bm{u}_M}/{\bm{u}_M^{\top}  \bm{R}^{-1} \bm{u}_M }$ contains at least one negative element. Let $d_3$ be the maximum value among all the negative elements of ${\bm{R}^{-1} \bm{u}_M}/{\bm{u}_M^{\top}  \bm{R}^{-1} \bm{u}_M }$.
Noticing that $\sum_{n=1}^{N_{\rm{hef}}+1} \bm{v}^{(n)}/ (N_{\rm{hef}}+1) \geq \bm{0}$, we see that  
if we select $ \epsilon_2< -d_3$, event $A_3$ is always false.
Therefore, by selecting a small enough  $\epsilon_2$, we have both
\begin{align}
\lim_{e_0 \rightarrow \infty} \mathbb{P} (A_1 \cap A_3) \leq \lim_{e_0 \rightarrow \infty} \mathbb{P} (A_3)=0,\\
\lim_{e_0 \rightarrow \infty} \mathbb{P} (A_1' \cap A_3) \leq \lim_{e_0 \rightarrow \infty} \mathbb{P} (A_3)=0.
\end{align}

(ii) We are left with the second scenario where either $\bm{R}^{-1}\bm{u}_M \geq \bm{0}$ or $\bm{R}^{-1}\bm{u}_M \leq \bm{0}$. In this scenario, we compute in Lemma~\ref{lm:dualoptimality} the optimal solution of problem (\ref{eq:max_dropping_rate}) analytically.
\begin{lemma}
Under the conditions that: (i) either $\bm{R}^{-1} \bm{u}_M\geq  \bm{0}$ or $\bm{R}^{-1} \bm{u}_M\leq  \bm{0}$,  and (ii) $(\bm{C}^{\top})^{-1} \bm{u}_M \geq \bm{0}$, the optimal solution of problem (\ref{eq:max_dropping_rate}) is
\begin{empheq}[left=\empheqlbrace]{align}
\label{eq:def_vstar} \bar{\bm{v}}^* &=\bm{R}^{-1} \bm{u}_M/\bm{u}_M^{\top} \bm{R}^{-1} \bm{u}_M,\\
\label{eq:def_fstar} {{f}}^* &= 1/\bm{u}_M^{\top} \bm{R}^{-1} \bm{u}_M.
\end{empheq}
\label{lm:dualoptimality}
\end{lemma}
\begin{IEEEproof}
The general idea of the proof is to use the duality properties of linear programmings. When either $\bm{R}^{-1}  \bm{u}_M\geq \bm{0}$ or $\bm{R}^{-1}  \bm{u}_M\leq \bm{0}$, $\bar{\bm{v}}=\bm{R}^{-1} \bm{u}_M /\bm{u}_M^{\top} \bm{R}^{-1} \bm{u}_M$ is a feasible solution of problem (\ref{eq:max_dropping_rate}), whose corresponding objective value is $1/\bm{u}_M^{\top} \bm{R}^{-1} \bm{u}_M$. Because $f^*$ is the optimal objective value of (\ref{eq:max_dropping_rate}), we have $1/\bm{u}_M^{\top} \bm{R}^{-1} \bm{u}_M\geq f^*$. In the following, we will show $f^* \geq 1/\bm{u}_M^{\top} \bm{R}^{-1} \bm{u}_M$.

The dual problem of (\ref{eq:max_dropping_rate}) is written as:
\begin{equation}
\begin{aligned}
&\max_{w,\bm{\lambda}}
& & w\\
& \text{s.t.} & & \bm{u}_M^\top  \bm{\lambda}=1,\\
&&&\bm{R}^\top  \bm{\lambda} \geq w \bm{u}_M,\\
&&& \bm{\lambda}\geq \bm{0}.
\end{aligned}
\label{eq:max_dropping_rate_dual}
\end{equation}
Because of the Sherman-Woodbury-Morrison formula and because $\bm{R}=\bm{C}-\bar{\bm{s}} \cdot \bm{u}_M^{\top}$, we have that
\[
(\bm{R}^\top)^{-1}\bm{u}_M= (\bm{C}^\top)^{-1} \bm{u}_M /(1-\bm{u}_M^{\top} \bm{C}^{-1} \bar{\bm{s}}).
\]
Because $(\bm{C}^{\top})^{-1} \bm{u}_M \geq \bm{0}$, we either have $(\bm{R}^{\top})^{-1}\bm{u}_M \geq \bm{0}$ or $(\bm{R}^{\top})^{-1}\bm{u}_M \leq \bm{0}$, depending on the difference between $\bm{u}_M^{\top} \bm{C}^{-1} \bar{\bm{s}}$ and $1$.  In both cases, we have a feasible solution of the dual problem
\[
\left\{
\begin{aligned}
\bm{\lambda}&=\frac{(\bm{R}^\top)^{-1} \bm{u}_M}{\bm{u}_M^\top (\bm{R}^\top)^{-1} \bm{u}_M},\\
w &= \frac{1}{\bm{u}_M^\top  (\bm{R}^\top)^{-1} \bm{u}_M}.
\end{aligned}
\right.
\]
Consequently, the objective value $ {1}/\left(\bm{u}_M^\top  (\bm{R}^\top)^{-1} \bm{u}_M\right)$ reached by this feasible solution of the dual problem provides a lower bound of the objective value for the original problem (\ref{eq:max_dropping_rate}), that is, $f^* \geq  {1}/\bm{u}_M^\top  (\bm{R}^\top)^{-1} \bm{u}_M$.
By noticing that 
\begin{align*}
1/{\bm{u}_M^\top\cdot (\bm{R}^\top)^{-1} \cdot \bm{u}_M}={1}/{\bm{u}_M^\top \cdot (\bm{R})^{-1} \cdot \bm{u}_M},
\end{align*}
we see that $f^*=1/\bm{u}_M^{\top} \bm{R}^{-1} \bm{u}_M$. The solution that attains this optimal objective value is $\bm{\bar{v}}^*= \bm{R}^{-1} \bm{u}_M/\bm{u}_M^{\top} \bm{R}^{-1} \bm{u}_M$.
\end{IEEEproof}

When event $A_3$ is true, Using (\ref{eq:relationhmem}), we see that
\begin{align}
\nonumber & h_{l^*}^{(N_{\rm{hef}}+1)}-h_{l^*}^{(0)}  \\
\nonumber & \qquad=  e_{l^*}^{(N_{\rm{hef}}+1)}- e_0+ \tau \sum_{n=1}^{N_{\rm{hef}}+1} (\bm{R} \bm{v}^{(n)})_{l^*} \\
\nonumber & \qquad <  -e_0 + \tau \sum_{n=1}^{N_{\rm{hef}}+1} (\bm{R} (\bm{v}^{(n)} - \bm{\bar{v}}^* ))_{l^*}+ \tau \sum_{n=1}^{N_{\rm{hef}}+1} (\bm{R} \bm{\bar{v}}^*)_{l^*} \\
& \qquad <  -e_0+ \tau (N_{\rm{hef}}+1) \cdot M \epsilon_2 \|\bm{R}\|_{\infty} +\tau  (N_{\rm{hef}}+1)\cdot f^*,
\label{eq:hlhefgelsmd}
\end{align}
where the inequality on the third line is because $e_{l^*}^{(N_{\rm{hef}}+1)}<0$, and the inequality on the fourth line follows from (\ref{eq:def_eventa3}) and $\bm{R} \bm{\bar{v}}^* \leq f^* \bm{u}_M$. We will use (\ref{eq:hlhefgelsmd}) in both the proof for points $2A$  and $2B$. 

1) We first look at point $2A$ and consider the event $A_1' \cap A_3$.

We denote by
\[
d_4= -f^*-  M\epsilon_2 \|\bm{R}\|_{\infty},
\]
which is positive when we select $\epsilon_2<-f^*/M\|\bm{R}\|_{\infty}$.

Plugging the definition of $d_4$ into (\ref{eq:hlhefgelsmd}) and using Corollary~\ref{coro:wsnazu}, we have
\begin{align*}
& \mathbb{P}(h_{l^*}^{(N_{\rm{hef}}+1)}-h_{l^*}^{(0)} < -e_0- \tau d_4 (N_{\rm{hef}}+1)) \\
& \qquad \leq  \exp \left( \frac{-(N_{\rm{hef}}+1)d_4^2}{2 S^2}\right) \exp \left(-\frac{e_0  d_4}{\tau S^2} \right) \\
& \qquad \leq  \exp \left(-\frac{e_0  d_4}{\tau S^2} \right).
\end{align*}
Using the union bound and using $N_{\rm{hef}} \leq Ke_0$, we have
\begin{align*}
\nonumber & \mathbb{P}(A_1' \cap A_3 ) \\
\nonumber & \qquad \leq   \sum_{N_{\rm{hef}}=1}^{Ke_0} \mathbb{P}\left(h_{l^*}^{(N_{\rm{hef}}+1)}-h_{l^*}^{(0)} < -e_0-\tau d_4 (N_{\rm{hef}}+1)\right) \\
& \qquad \leq  Ke_0 \exp \left(-\frac{e_0  d_4}{\tau S^2} \right).
\end{align*}
Taking $e_0 \rightarrow \infty$, we see that $\lim_{e_0 \rightarrow \infty}\mathbb{P} (A_1' \cap A_3 ) =0$.

2) Then, we show point $2B$ by considering the event $A_1 \cap A_3$. Through (\ref{eq:hlhefgelsmd}), we have
\begin{align}
\nonumber  & h_{l^*}^{(N_{\rm{hef}}+1)}-h_{l^*}^{(0)}  \\
\nonumber & \quad <  -e_0+ (M \epsilon_2 \|\bm{R}\|_{\infty}  + f^* )\tau (N_{\rm{hef}}+1) \\
\label{eq:sdfwerfdguidug}
 & \quad \leq -e_0+ (M\epsilon_2 \|\bm{R}\|_{\infty}+f^*) ((1-\delta)e_0/ f^*+\tau)\\
\nonumber & \quad = -\left( \delta - \frac{M\epsilon_2 (1-\delta) \|\bm{R}\|_{\infty}}{f^*}\right) e_0 +\tau (M\epsilon_2 \|\bm{R}\|_{\infty}+f^*),
\end{align}
where the inequality on the third line is because $N_{\rm{hef}} \leq (1-\delta)e_0/ (\tau f^*)$. 
We define a small constant $\zeta>0$ and see that $\tau (M\epsilon_2 \|\bm{R}\|_{\infty}+f^*) \leq \zeta e_0$ when $e_0$ is large. Define
\[
d_5=\delta - M\epsilon_2(1-\delta) \|\bm{R}\|_{\infty}/f^*- \zeta,
\]
which is positive when we set $\epsilon_2 < (\delta-\zeta)  f^*/ M(1-\delta)\|\bm{R}\|_{\infty}$.

Plugging $d_5$ into (\ref{eq:sdfwerfdguidug}) and using Corollary~\ref{coro:wsnazu},  
\begin{align*}
\nonumber \mathbb{P}(h_{l^*}^{(N_{\rm{hef}}+1)}-h_{l^*}^{(0)}<- d_5 e_0) \leq & \exp \left(-\frac{d_5^2 e_0^2}{2 \tau^2 S^2} \right).
\end{align*}
Using the union bound and using $N_{\rm{hef}} \leq (1-\delta)e_0/(\tau f^*)$, we have
\begin{align*}
\nonumber & \mathbb{P}(A_1 \cap A_3 ) \\
\nonumber & \qquad \leq   \sum_{N_{\rm{hef}}=1}^{(1-\delta)e_0/(\tau f^*)} \mathbb{P}\left(h_{l^*}^{(N_{\rm{hef}}+1)}-h_{l^*}^{(0)} < - d_5 e_0\right) \\
& \qquad \leq  \frac{(1-\delta)e_0}{\tau f^*} \exp \left(-\frac{d_5^2 e_0^2}{2 \tau^2 S^2} \right).
\end{align*}
Taking $e_0 \rightarrow \infty$, we see that $\lim_{e_0 \rightarrow \infty}\mathbb{P} (A_1 \cap A_3 ) =0$.

\subsection{Proof for point $3A$ and point $3B$}
The proofs for $3A$ and point $3B$ are identical. Therefore, we only show the proof for $3A$ here.
We define the event $A_5$ as
\begin{equation}
A_5=\left\{\frac{1}{N_{\rm{hef}}+1} \left\|\sum_{n=1}^{N_{\rm{hef}}+1} \bm{s}^{(n)}-\bm{\bar{s}}\right\|_{\infty}< \epsilon_4\right\},
\label{eq:defa5}
\end{equation}
where $\epsilon_4$ is a small constant.

We first show that event $A_1' \cap A_2 \cap A_5 \subseteq A_1' \cap A_3$.
Using (\ref{eq:sum_iterative_function}) and (\ref{eq:defR}), we calculate the difference between the available energy of all BSs at time $0$ and that at time $N_{\rm{hef}}+1$:
\[
\bm{e}^{(N_{\rm{hef}}+1)}=\bm{e}^{(0)}-\sum_{n=1}^{N_{\rm{hef}}+1} \bm{R} \bm{v}^{(n)} +\sum_{n=1}^{N_{\rm{hef}}+1} \left( \bm{s}^{(n)}- \bar{\bm{s}}\right),
\]
which is equivalent to 
\begin{align}
\nonumber & \frac{1}{N_{\rm{hef}}+1}\sum_{n=1}^{N_{\rm{hef}}+1} \bm{v}^{(n)}- \frac{1}{N_{\rm{hef}}+1}\bm{R}^{-1}(\bm{e}^{(0)}-\bm{e}^{(N_{\rm{hef}}+1)} ) \\
& \qquad =(\bm{R})^{-1}\frac{\sum_{n=1}^{N_{\rm{hef}}+1} \left( \bm{s}^{(n)}- \bar{\bm{s}}\right) }{N_{\rm{hef}}+1}.
\label{eq:vnconvergstes1}
\end{align}
Multiplying $\bm{u}_M^\top$ on both sides and using  $\bm{u}_M^{\top} \cdot \sum_{n=1}^{N_{\rm{hef}}+1} \bm{v}^{(n)}=N_{\rm{hef}+1}+1$, we see that
\begin{equation}
1- \bm{u}_M^{\top} \bm{R}^{-1}\frac{\bm{e}^{(0)}-\bm{e}^{(N_{\rm{hef}}+1)}}{N_{\rm{hef}}+1}= \bm{u}_M^{\top} \bm{R}^{-1}\frac{  \sum_{n=1}^{N_{\rm{hef}}+1} \left( \bm{s}^{(n)}- \bar{\bm{s}}\right)}{N_{\rm{hef}}+1}.
\label{eq:sdftranformfd}
\end{equation}
Taking the infinite norm on both sides and knowing that event $A_5$ (\ref{eq:defa5}) holds true, we transform (\ref{eq:sdftranformfd}) into 
\begin{equation}
\left| 1- \frac{ \bm{u}_M^{\top} \bm{R}^{-1}(\bm{e}^{(0)}-\bm{e}^{(N_{\rm{hef}}+1)})}{N_{\rm{hef}}+1} \right|< M \|\bm{u}_M^\top \bm{R}^{-1}\|_{\infty}\epsilon_4.
\label{eq:eoplefdotnu0}
\end{equation}
Then, since $\bm{e}^{(0)}= e_0 \bm{u}_M$, (\ref{eq:eoplefdotnu0}) becomes 
\begin{align}
\nonumber \left|  1-\frac{e_0 \bm{u}_M^{\top} \bm{R}^{-1} \bm{u}_M}{N_{\rm{hef}}+1}  \right| &< M \|\bm{u}_M^\top \bm{R}^{-1}\|_{\infty}\epsilon_4+ \left|  \frac{ \bm{u}_M^{\top} \bm{R}^{-1} \bm{e}^{(N_{\rm{hef}}+1)}}{N_{\rm{hef}}+1} \right| \\
\nonumber & < M\|\bm{u}_M^\top \bm{R}^{-1}\|_{\infty}  \left( \epsilon_4 +\frac{\|\bm{e}^{(N_{\rm{hef}}+1)}\|_{\infty}} {N_{\rm{hef}}+1 } \right) \\
\nonumber & < M\|\bm{u}_M^\top \bm{R}^{-1}\|_{\infty} \left( \epsilon_4 +\epsilon_1 d_2\right) \\
& =\epsilon_5,
\label{eq:eoplefdotnu}
\end{align}
where the third inequality holds because $N_{\rm{hef}}+1 \geq e_0/d_2$ and event $A_2$ holding true gives $\|\bm{e}^{(N_{\rm{hef}}+1)}\|_{\infty}\leq \epsilon_1 e_0$, and the fourth inequality comes from the definition
\[
\epsilon_5=  M\|\bm{u}_M^\top \bm{R}^{-1}\|_{\infty} \cdot \left( \epsilon_4 + \epsilon_1 d_2 \right).
\]
We recast (\ref{eq:eoplefdotnu}) as 
\begin{equation}
\frac{1-\epsilon_5}{\bm{u}_M^{\top} \bm{R}^{-1} \bm{u}_M}< \frac{e_0}{ N_{\rm{hef}}+1}<\frac{1+\epsilon_5}{\bm{u}_M^{\top} \bm{R}^{-1} \bm{u}_M}.
\label{eq:erjioojijsdjf}
\end{equation}
We transform (\ref{eq:vnconvergstes1}) into
\begin{align}
\nonumber &\left\| \frac{1}{N_{\rm{hef}}+1}\sum_{n=1}^{N_{\rm{hef}}+1} \bm{v}^{(n)} - \frac{\bm{R}^{-1} \bm{e}^{(0)}  }{N_{\rm{hef}} +1} \right\|_{\infty}  \\
\nonumber & \qquad \leq  \|\bm{R}^{-1}\|_{\infty} \cdot \left\|\frac{-\bm{e}^{(N_{\rm{hef}}+1)}+\sum_{n=1}^{N_{\rm{hef}}+1} (\bm{s}^{(n)}- \bar{\bm{s}} ) }{N_{\rm{hef}}+1} \right\|_{\infty}\\
\nonumber & \qquad \leq  \|\bm{R}^{-1}\|_{\infty} \left(\left\| \frac{\bm{e}^{(N_{\rm{hef}}+1)}}{N_{\rm{hef}}+1}\right\|_{\infty}+ \sum_{n=1}^{N_{\rm{hef}+1}}  \frac{\|\bm{s}^{(n)}- \bar{\bm{s}}\|_{\infty}}{N_{\rm{hef}}+1}  \right)\\  
\nonumber & \qquad \leq \|\bm{R}^{-1}\|_{\infty}  \left(  \frac{\epsilon_1 e_0}{e_0/d_2}+\epsilon_4 \right)\\
& \qquad \leq \|\bm{R}^{-1}\|_{\infty}  (\epsilon_4 + \epsilon_1 d_2),
\label{eq:fsadfjkkjfgjioron}
\end{align}
where the third inequality is because $\|\bm{e}^{(N_{\rm{hef}}+1)}\|_{\infty}< \epsilon_1 e_0 $, because $N_{\rm{hef}}+1 \geq e_0/d_2$ and because of (\ref{eq:defa5}). 
Plugging (\ref{eq:erjioojijsdjf}) into (\ref{eq:fsadfjkkjfgjioron}), we have
\begin{equation}
\left| \frac{1}{N_{\rm{hef}}+1}\sum_{n=1}^{N_{\rm{hef}}+1} \bm{v}^{(n)} - \frac{\bm{R}^{-1} \bm{u}_M  }{\bm{u}_M^{\top}  \bm{R}^{-1} \bm{u}_M } \right|<\epsilon_2\bm{u}_M,
\label{eq:vmagsrmsdj}
\end{equation}
where  we set
\[
\epsilon_2= \left\|\bm{R}^{-1} \right\|_{\infty} (\epsilon_4+ \epsilon_1 d_2 )+ \frac{\epsilon_5 \|\bm{R}^{-1} \bm{u}_M\|_{\infty}}{\bm{u}_M^\top \bm{R}^{-1} \bm{u}_M}.
\]
Then, (\ref{eq:vmagsrmsdj}) shows that $A_3$ occurs, and therefore $A_1'\cap {A}_2 \cap A_5  \subseteq A_1' \cap A_3$ if we choose $\epsilon_1, \epsilon_2, \epsilon_4, \epsilon_5$ properly. 
Hence, if $\lim_{e_0 \rightarrow \infty} \mathbb{P} (A_1' \cap A_3)=0$,
\[
\lim_{e_0 \rightarrow \infty} \mathbb{P}(A_1'\cap {A}_2 \cap A_5)=0.
\]
Moreover, because of the rule of additions of probabilities,
\begin{align}
\nonumber \mathbb{P} (A_1' \cap {A}_2) + \mathbb{P} (A_5)  & =  \mathbb{P} (A_1'\cap {A}_2 \cap A_5) +\mathbb{P} ((A_1'\cap {A}_2) \cup A_5)\\
& \leq  \mathbb{P} (A_1'\cap {A}_2 \cap A_5) +1.
\end{align}
Because $\lim_{e_0 \rightarrow \infty} \mathbb{P} (A_5)=1$ holds true using the weak law of large numbers with (\ref{eq:recharge_rate_constant_expectation}) and (\ref{eq:recharge_rate_boundedness}), we have $\lim_{e_0 \rightarrow \infty} \mathbb{P} (A_1'\cap {A}_2) =0$.

\end{document}